\documentclass[10pt,aps,pra,twocolumn,superscriptaddress,floatfix,longbibliography]{revtex4-1}

\usepackage{amssymb}
\usepackage{amsmath}
\usepackage{graphicx}
\usepackage{dcolumn}
\usepackage{xcolor}
\usepackage{bm}
\usepackage{enumitem}
\usepackage{amsfonts}%
\usepackage{bbold}
\usepackage[colorlinks=true, allcolors=blue]{hyperref}
\usepackage{amsmath}
\DeclareMathOperator{\Tr}{Tr}

\providecommand{\U}[1]{\protect\rule{.1in}{.1in}}

\begin{document}

\title{Multi-channel fluctuating field approach to competing instabilities \\in interacting electronic systems}

\author{E. Linn\'er}
\affiliation{CPHT, CNRS, Ecole Polytechnique, Institut Polytechnique de Paris, F-91128 Palaiseau, France}

\author{A. I. Lichtenstein}
\affiliation{I. Institute of Theoretical Physics, University of Hamburg, Jungiusstrasse 9, 20355 Hamburg, Germany}
\affiliation{European X-Ray Free-Electron Laser Facility, Holzkoppel 4, 22869 Schenefeld, Germany}
\affiliation{The Hamburg Centre for Ultrafast Imaging, Luruper Chaussee 149, 22761 Hamburg, Germany}

\author{S. Biermann}
\affiliation{CPHT, CNRS, Ecole Polytechnique, Institut Polytechnique de Paris, F-91128 Palaiseau, France}
\affiliation{Coll\`ege de France, 11 place Marcelin Berthelot, 75005 Paris, France}
\affiliation{Department of Physics, Division of Mathematical Physics, Lund University, Professorsgatan 1, 22363 Lund, Sweden}
\affiliation{European Theoretical Spectroscopy Facility, 91128 Palaiseau, France}

\author{E. A. Stepanov}
\affiliation{CPHT, CNRS, Ecole Polytechnique, Institut Polytechnique de Paris, F-91128 Palaiseau, France}

\begin{abstract}
Systems with strong electronic Coulomb correlations often
display rich phase diagrams exhibiting different ordered
phases involving spin, charge, or orbital degrees of freedom.
The theoretical description of the interplay of the
corresponding collective fluctuations giving rise to this
phenomenology remains however a tremendous challenge.
Here, we introduce a multi-channel extension of the recently 
developed fluctuating field approach to competing collective fluctuations in correlated electron systems. 
The method is based on a variational optimization of a trial action that explicitly contains the order parameters of the leading fluctuation channels.
It gives direct access to the free energy of the system, 
facilitating the distinction between stable and meta-stable phases of the system.
We apply our approach to the extended Hubbard model in the
weak to intermediate coupling regime where we find it to
capture the interplay of competing charge density wave and antiferromagnetic fluctuations with qualitative agreement with more computationally expensive methods.
The multi-channel fluctuation field approach thus offers a promising
new route for a numerically cheap treatment of the interplay between collective fluctuations in large systems.
\end{abstract}

\maketitle

\section{Introduction}
A hallmark of materials with strong electronic Coulomb correlations are their typically extremely rich phase diagrams, exhibiting various kinds of ordering phenomena. These result from competing instabilities involving e.g. charge, spin, orbital or pairing fluctuations. The theoretical description of these collective phenomena remains a challenging issue of computational complexity~\cite{jres.045.026,PhysRevLett.56.2521} as well as conceptual difficulty, e.g. the explicit breaking of symmetries~\cite{10.1143/PTP.112.943,RevModPhys.35.496}.
In this sense, the interplay of competing electronic fluctuations constitutes a roadblock to the understanding of the complex phase diagrams of a wide range of material systems.
Constructing simplified methods to study interplaying collective fluctuations is thus of crucial importance.

The extended Hubbard model~\cite{rspa.1963.0204,PhysRevLett.10.159,10.1143/PTP.30.275,rspa.1964.0190} provides a suitable framework for investigating the interplay between collective electronic fluctuations.
The physics of this model is determined by the competition between the local $U$ and the non-local $V$ Coulomb interactions. 
A repulsive $U$ stabilizes collective spin fluctuations~\cite{PhysRev.157.295}, which may compete with charge fluctuations driven by a strong repulsive $V$~\cite{PhysRevB.3.2662, Vonsovsky_1979}. 
The earliest considerations of the extended Hubbard model were already implicit in the initial work of J. Hubbard in 1963~\cite{rspa.1963.0204}. 
However, the first studies of the model occurred in the 1970's, with studies of the strong~\cite{PhysRevB.3.2662, PhysRevB.14.2989} and weak coupling limits of the half-filled one-dimensional (1D) chain~\cite{doi:10.1080/00018737900101375, emery1979}.
Together with an access to the intermediate coupling regime by early numerical exact diagonalization (ED) and lattice Monte Carlo calculations~\cite{PhysRevB.29.5096, PhysRevLett.53.2327}, the phase diagram of the 1D extended Hubbard model was predicted to be composed of regions of strong charge density wave (CDW) and antiferromagnetic (AFM) fluctuations, with a CDW-AFM transition occurring in the vicinity of ${U=2V}$.
The transition was later discovered to be modified in the weak coupling limit by an intermediate bond-order wave (BOW) state~\cite{doi:10.1143/JPSJ.68.3123, PhysRevB.61.16377}. 

Extensive studies have been conducted on the extended Hubbard model for elucidating the interplay between collective charge and spin fluctuations~{\cite{PhysRevB.14.2989, emery1979, PhysRevB.29.5089, PhysRevB.29.5096, PhysRevLett.53.2327, PhysRevB.39.9397, PhysRevB.42.465, PhysRevB.48.7140, PhysRevB.70.235107, PhysRevB.74.035113, PhysRevB.99.245146, PhysRevB.100.075108, stepanov2021coexisting, Vandelli}}.
Considerable insight has been acquired for the extended Hubbard model on a two-dimensional square lattice at half-filling with nearest-neighbour interaction $V$~\cite{PhysRevB.39.9397, PhysRevB.42.465, PhysRevB.48.7140, PhysRevB.70.235107, PhysRevB.74.035113, PhysRevB.87.125149, PhysRevB.90.235105, PhysRevB.95.115149, van_Loon_2018, PhysRevB.99.115112, PhysRevB.99.245146, PhysRevB.100.075108, PhysRevB.102.195109, PhysRevB.104.085129, PhysRevB.94.205110, PhysRevB.102.195109, PhysRevB.99.115124}, which we study in the current work.
It has been found that this model displays a phase diagram similar to the one-dimensional counterpart, besides the apparent lack of an intermediate BOW phase.
In particular, the system reveals a checker-board CDW pattern which interplays with strong AFM fluctuations in the vicinity of a CDW-AFM transition line ${U=4V}$~\cite{PhysRevB.39.9397}.
In a recent work~\cite{PhysRevB.99.245146} based on the dynamical cluster approximation (DCA)~\cite{PhysRevB.58.R7475, PhysRevB.61.12739, PhysRevB.65.153102}, the competition near the transition line has been shown to induce a coexistence region of charge- and spin-ordered states.

By the Mermin-Wagner theorem~\cite{PhysRev.158.383, PhysRevLett.17.1133, PhysRev.171.513}, magnetic ordering at finite temperatures is excluded in a broad class of one- and two-dimensional systems, including the extended Hubbard model, due to the continuous nature of the underlying symmetry.
Thus, the regime of strong collective AFM fluctuations is strictly speaking not a phase.
However, in our current work the ``AFM phase'' will refer to a slightly broader definition of short-range AFM ordering, which transforms to a true phase for a quasi-two-dimensional system.
In contrast, the discrete symmetry of the CDW allows for a true phase transition.
In addition, technically speaking, in the present work, we are performing calculations for finite systems, where long-range fluctuations are eventually cut off, so neither the AFM or CDW state are strictly speaking phases.
Nevertheless, in the following, we will refer to both states as phases, since we are interested in the interplay of the competing fluctuations corresponding to these orderings.
Our conclusions should thus be understood as applying either to finite systems replacing the notion of phase by "state dominated by the respective fluctuations" or to a quasi-two-dimensional system in the thermodynamic limit.

Limitations in the treatment of competing collective fluctuations arise in the currently available approaches employed for studying quantum lattice systems.
Numerically exact methods, such as exact diagonalization (ED)~\cite{jres.045.026} and lattice Monte Carlo~\cite{PhysRevLett.56.2521} have studied the interplay between $U$ and $V$~\cite{PhysRevB.29.5096,PhysRevLett.53.2327,PhysRevB.39.9397,PhysRevB.42.465} but are restricted to small system sizes and thus cannot address long-range collective fluctuations.
The same problem is also inherent in cluster extensions of the dynamical mean-field theory (DMFT)~\cite{PhysRevB.62.R9283, PhysRevLett.87.186401, RevModPhys.77.1027, doi:10.1063/1.2199446, RevModPhys.78.865, PhysRevB.94.125133}, such as, e.g., DCA~\cite{PhysRevB.58.R7475, PhysRevB.61.12739, PhysRevB.65.153102}.
Diagrammatic methods based on the parquet approximation~\cite{osti_4338008, 1.1724313, 1.1704062, BICKERS1989206, PhysRevB.43.8044, Bickers2004} allow one to account for the interplay between charge and spin fluctuations~\cite{PhysRevB.100.075108} originating from the two-particle vertex functions in an unbiased and powerful fashion. 
These vertices are incorporated with full momentum- and frequency-dependence, and the approach is thus computationally very expensive, which severally limits its applicability.
Advanced diagrammatic extensions of DMFT~\cite{RevModPhys.90.025003} are able to describe long-range fluctuations simultaneously in different instability channels.
In the presence of the non-local interaction $V$ this can be done within the dual boson theory~\cite{PhysRevB.90.235135, PhysRevB.93.045107, PhysRevB.94.205110, PhysRevB.100.165128,PhysRevB.102.195109}, the dynamical vertex approximation (D$\Gamma$A)~\cite{PhysRevB.95.115107, doi:10.7566/JPSJ.87.041004}, the triply irreducible local expansion (TRILEX) method~\cite{PhysRevB.97.155145}, or the dual TRILEX (\mbox{D-TRILEX}) approach~\cite{stepanov2021coexisting, arxiv.2204.06426, Vandelli}.
However, these fluctuations are usually treated in a ladder-like approximation, where different instability channels affect each other only indirectly via self-consistent renormalization of single- and two-particle quantities.

Current approaches to quantum lattice systems that are able to capture competing collective fluctuations are too complicated for broad usage.
In this work, we develop a multi-channel generalisation of the fluctuating field (FF) approach that allows us to incorporate multiple collective fluctuation channels and their interplay in a numerically cheap way without explicitly breaking the symmetry of the model. 
The FF method was originally introduced for the study of spin fluctuations in the classical Ising plaquettes~\cite{PhysRevE.97.052120} and was further developed for single- and multi-mode treatment of collective spin fluctuations in the Hubbard model~\cite{PhysRevB.102.224423, PhysRevB.105.035118, s10948-022-06303-8}. 
We employ the proposed multi-channel fluctuating field (MCFF) approach to study the interplay between CDW and AFM fluctuations in the extended Hubbard model on a half-filled square lattice with a repulsive on-site $U$ and nearest-neighbour $V$ interactions. 
We show that the MCFF approach predicts results for the CDW and AFM phase boundaries in qualitative agreement with more elaborate numerical methods.
Furthermore, it allows to model competing collective fluctuations for large system sizes near the thermodynamic limit. 
In addition, the method is able to distinguish between stable and meta-stable collective fluctuations. 
For this reason, the MCFF approach allows us to capture the true ground state of the coexistence region of CDW and AFM fluctuation that was obtained in Ref.~\cite{PhysRevB.99.245146} on the basis of DCA calculations.

\section{Model}
For simplicity, our considerations are limited to a single-band extended Hubbard model. 
However, we note that our approach can be straightforwardly generalised to more complex single- and multi-band quantum lattice systems.
The Hamiltonian of the extended Hubbard model has the following form:
\begin{align}
\hat{H} = 
- t \sum_{\langle i,j \rangle,\sigma} \hat{c}^\dagger_{i \sigma} \hat{c}^{\phantom{\dagger}}_{j \sigma} 
+ U \sum_{i} \hat{n}_{i \uparrow} \hat{n}_{i \downarrow} 
+ \frac{V}{2} \sum_{\langle i,j\rangle,\sigma\sigma'} \hat{n}_{i \sigma} \hat{n}_{j \sigma'}.
\label{Eq:EHubbard_Ham}
\end{align}
In this expression, $\hat{c}_{i\sigma}^{(\dagger)}$ operators correspond to annihilation (creation) of electrons, where the subscripts denote the position $i$ and spin projection ${\sigma \in \{ \uparrow, \downarrow \}}$. 
Our system is modelled by the hopping $t$ between nearest-neighbor sites $\langle i,j \rangle$ on a two-dimensional square lattice. 
The Coulomb interaction between electronic densities ${\hat{n}^{\phantom{\dagger}}_{i \sigma} = \hat{c}^\dagger_{i \sigma} \hat{c}^{\phantom{\dagger}}_{i \sigma}}$ contains the on-site $U$ and the nearest-neighbor $V$ components.

\begin{figure}[t!]
\includegraphics[width=1.\linewidth]{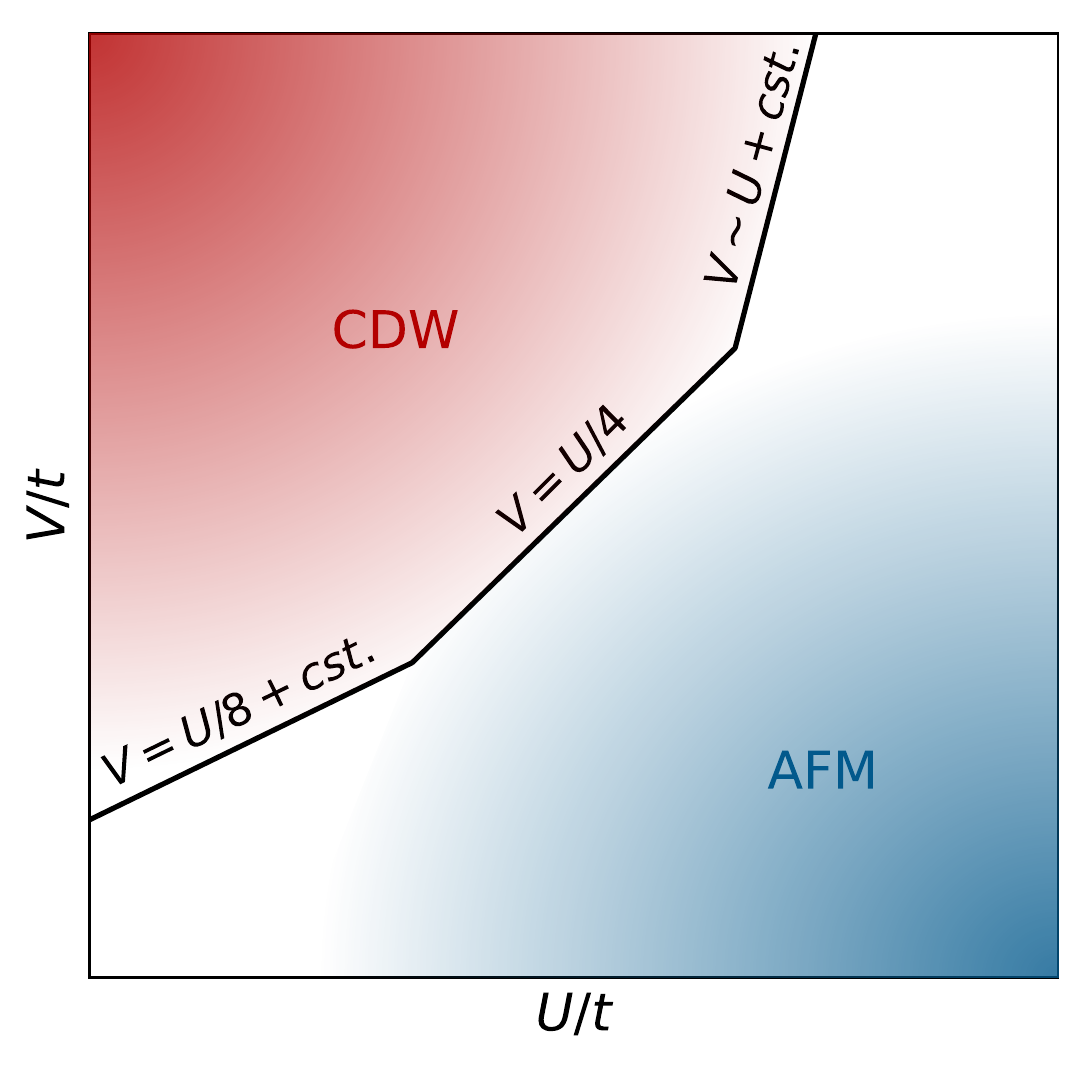}
\caption{Sketch of the phase diagram of the quasi-two-dimensional half-filled extended Hubbard model with repulsive interactions $U$ and $V$ at low but finite temperature. Beyond a critical local interactions, a regime of dominant antiferromagnetic (AFM) fluctuations is expected, while strong non-local interactions drive the system into a charge density wave (CDW) phase. At low $U$ and $V$, the orderings give away for a homogeneous paramagnetic (PM) phase.
The schematic phase boundaries of the CDW phase is determined by the asymptotic expressions 
${V=U/8 +\text{cst.}}$ at weak coupling~\cite{PhysRevB.99.115112}, ${V=U/4}$ at intermediate coupling~\cite{PhysRevB.39.9397,PhysRevB.42.465} and ${V\sim U + \text{cst.}}$ at strong coupling~\cite{PhysRevB.87.125149,PhysRevB.90.235135,PhysRevB.95.245130,PhysRevB.94.205110,PhysRevB.102.195109}.
At weak to intermediate coupling, the AFM regime extrapolates from a critical $U$ at vanishing $V$ to the {$V=U/4$} phase boundary~\cite{PhysRevB.99.245146}.
\label{figure1}}
\end{figure}

The extended Hubbard model~\eqref{Eq:EHubbard_Ham} displays two symmetries of fundamental importance for our considerations: a continuous SU(2) symmetry associated with spin degrees of freedom and a discrete particle-hole symmetry related to charge degrees of freedom.
To facilitate our later treatments, we include a sketch of the finite temperature $U,V$ phase diagram of the extended Hubbard model on the two-dimensional square lattice in Fig.~\ref{figure1}.
Within the sketch, we denote the regime of strong CDW fluctuations (red gradient), with asymptotics of the CDW phase boundary highlighted, and the regime of strong AFM fluctuations (blue gradient).
The CDW phase boundary occurs along ${V=U/8 +\text{cst.}}$ at weak coupling~\cite{PhysRevB.99.115112}, which transforms to ${V=U/4}$ at intermediate coupling~\cite{PhysRevB.39.9397}, followed by ${V\sim U + \text{cst.}}$ at strong coupling~\cite{PhysRevB.87.125149,PhysRevB.90.235135,PhysRevB.95.245130,PhysRevB.94.205110,PhysRevB.102.195109}.
At weak coupling the AFM phase boundary starts at a critical $U$, which further extends to the {$V=U/4$} phase boundary at intermediate coupling~\cite{PhysRevB.99.245146}.
We restrict our consideration to the weak to intermediate coupling regime, with the strong coupling regime being outside the scope of the current work.

The MCFF approach to be introduced in the next section is based on a variational principle conveniently formulated within the action formalism. Thus, it is suitable to rewrite the extended Hubbard model~\eqref{Eq:EHubbard_Ham} in the form of the action:
\begin{align}
\mathcal{S} = &- \frac{1}{\beta N}\sum_{\bf{k},\nu ,\sigma} c_{\bf{k}\nu\sigma}^* \mathcal{G}_{\bf{k}\nu}^{-1} c^{\phantom{*}}_{\bf{k}\nu\sigma} 
+ \frac{U}{\beta N} \sum_{\bf{q},\omega} \rho_{\bf{q}\omega\uparrow} \rho_{-\bf{q},-\omega\downarrow} \nonumber \\
&+ \frac{1}{2 \beta N}\sum_{\bf{q},\omega,\sigma\sigma'}V_{\bf{q}} \rho_{\bf{q}\omega\sigma}\rho_{-\bf{q},-\omega\sigma'},
\label{Eq:EHubbard_Action}
\end{align}
with the inverse temperature $\beta$ and number of sites $N$. 
Grassmann variables $c^{(*)}$ correspond to the annihilation (creation) of electrons, where the subscripts denote the momentum $\bf{k}$ and fermionic Matsubara frequency $\nu$. 
The inverse of the bare (non-interacting) Green's function is defined as ${\mathcal{G}^{-1}_{\bf{k}\nu} = i\nu + \mu - \epsilon_{\bf{k}}}$, where $\mu$ is the chemical potential and ${\epsilon_{\bf{k}}=-2t(\cos{k_x} + \cos{k_y})}$ is the dispersion relation for the nearest-neighbor hopping on a two-dimensional square lattice. 
For convenience, the interaction parts of the action~\eqref{Eq:EHubbard_Action} are written in terms of the shifted densities ${\rho_{\bf{q}\omega\sigma} = n_{\bf{q}\omega\sigma} - \langle n_{\bf{q}\omega\sigma}\rangle\delta_{\bf{q},\bf{0}}\delta_{\omega,0}}$, where ${\bf q}$ and $\omega$ are the momentum and bosonic Matsubara frequency indices, respectively.
This choice of shift will be argued for in our later derivation.
In our considerations the momentum-space representation for the non-local interaction is following ${V_{\bf{q}} = 2V(\cos{q_x} + \cos{q_y})}$ as it is limited to only a nearest-neighbour interaction.

\section{Multi-channel fluctuating field method}
In this section we derive a multi-channel generalisation of the fluctuating field method that was originally introduced to address the fluctuations in a single (magnetic) channel~\cite{PhysRevE.97.052120, PhysRevB.102.224423, PhysRevB.105.035118, s10948-022-06303-8}.
We derive the MCFF method by utilizing a variational approach formulated in Ref.~\onlinecite{PhysRevB.102.224423}, which allows to incorporate the leading instabilities of the collective fluctuations.

\subsection{Definition of trial action}
We define a MC-FF trial action
\begin{align}
\mathcal{S}^{*} =  &- \frac{1}{\beta N}\sum_{\bf{k},\nu,\sigma} c_{\bf{k}\nu\sigma}^* \mathcal{G}_{\bf{k}\nu}^{-1} c^{\phantom{*}}_{\bf{k}\nu\sigma} \notag\\
&+ \sum_{\bf{Q},\varsigma} \left[ \phi_{\bf{Q}}^\varsigma \rho^\varsigma_{-\bf{Q}} - \frac{1}{2}\frac{\beta N}{J^\varsigma_{\bf{Q}}}\phi_{\bf{Q}}^\varsigma \phi_{-\bf{Q}}^\varsigma \right],
\label{Eq:EHubbard_FF_Action}
\end{align}
that explicitly considers sets of scalar charge (${\varsigma=c}$) and vector spin (${\varsigma = s \in \{x,y,z\}}$) fields $\phi^\varsigma_{\bf{Q}}$ coupled to the operators ${\rho^{\varsigma}_{\bf{Q}} = n^{\varsigma}_{\bf{Q}} - \langle n^{\varsigma}_{\bf{Q}}\rangle\delta^{\phantom{*}}_{\bf{Q},\bf{0}}}$ associated with the respective classical (${\omega=0}$) order parameters of interest.
Here
\begin{align}
n^\varsigma_{{\bf{Q}}} = \frac{1}{\beta{}N}\sum_{{\bf k}, \nu, \sigma\sigma'} c^{*}_{\bf{k+Q},\nu\sigma}\sigma^{\varsigma}_{\sigma\sigma'}c^{\phantom{*}}_{{\bf k}\nu\sigma'},
\end{align}
where ${\bf Q}$ is the ordering wave vector, $\sigma^{c}$ is the identity and $\sigma^{s}$ is the Pauli spin matrices.
The interaction part of the trial action~\eqref{Eq:EHubbard_FF_Action} contains a set of stiffness constants $J^{\varsigma}_{\bf{Q}}$ that will be determined.

\subsection{Integrating out fermionic degrees of freedom}
The trial action~\eqref{Eq:EHubbard_FF_Action} has a Gaussian form with respect to the Grassmann variables $c^{(*)}$ and classical fields $\phi^\varsigma$.
This allows one to obtain an effective action for either fermionic or classical degrees of freedom by analytically integrating out the other degrees of freedom. 
Integrating out the fermionic degrees of freedom, the effective action for the classical fields becomes:
\begin{align}
\mathcal{S}_{\phi} = &- \Tr \ln \left[ \mathcal{G}^{-1}_{{\bf k}\nu}\delta^{\phantom{*}}_{{\bf Q},0}\delta^{\phantom{*}}_{\sigma,\sigma'} - \sum_{\varsigma}\phi_{\bf{Q}}^\varsigma \sigma^{\varsigma}_{\sigma\sigma'} \right] \notag\\
&- \frac{1}{2}\sum_{\bf{Q},\varsigma}\frac{\beta N}{J^\varsigma_{\bf{Q}}}\phi_{\bf{Q}}^\varsigma \phi_{-\bf{Q}}^\varsigma.
\label{Eq:EHubbard_FF_Action_FluctuatingFields}
\end{align}
The trace is taken over the momenta ${\bf k, Q}$, frequency $\nu$, and spin ${\sigma, \sigma'}$ indices.
The effective action~\eqref{Eq:EHubbard_FF_Action_FluctuatingFields} depends on a small number of classical fields $\phi^{\varsigma}_{\bf Q}$.
For this reason, the phase diagram that captures the interplay between the different fluctuating fields can be studied by means of the free energy $\mathcal{F}_{\phi}$ corresponding to this action. 
Importantly, $\mathcal{F}_{\phi}$ non-perturbatively incorporates the fluctuations of the relevant order parameters $\rho_{\bf{Q}}^\varsigma$ by allowing the global minimum of $\mathcal{F}_{\phi}$ to shift away from ${\phi^\varsigma_{\bf Q}=0}$. 

\subsection{Determination of the stiffness parameters via a variational principle}
In order to determine $J^\varsigma_{\bf{Q}}$, we use the Peierls-Feynman-Bogoliubov variational principle~\cite{PhysRev.54.918,Bogolyubov:1958zv,feynman1972}, as previously employed for the single-mode FF method~\cite{PhysRevB.102.224423}. 
This variational principle allows one to construct a unique and unambiguous set of $J^\varsigma_{\bf{Q}}$ which minimizes the functional
\begin{align}
\mathcal{F}(J^\varsigma_{\bf{Q}}) & = \mathcal{F}_{c}(J^\varsigma_{\bf{Q}}) + \frac{1}{\beta N}\left\langle \mathcal{S} -\mathcal{S}_{c} \right\rangle_{{\cal S}_c}
\label{Eq:F_variational}
\end{align}
by varying $J^\varsigma_{\bf{Q}}$.
Here, $\langle \ldots \rangle_{{\cal S}_c}$ denotes the expectation value with respect to the effective fermionic action $\mathcal{S}_{c}$, corresponding to the trial action~\eqref{Eq:EHubbard_FF_Action} with the classical fields $\phi^{\varsigma}_{\bf Q}$ being integrated out:
\begin{align}
\mathcal{S}_{c} & = - \frac{1}{\beta N}\sum_{\bf{k},\nu,\sigma} c_{\bf{k}\nu\sigma}^* \mathcal{G}_{\bf{k}\nu}^{-1} c^{\phantom{*}}_{\bf{k}\nu\sigma} + \frac{1}{2} \sum_{\bf{Q},\varsigma}\frac{J^\varsigma_{\bf{Q}}}{\beta N}\rho_{\bf{Q}}^\varsigma \rho_{-\bf{Q}}^\varsigma.\label{Eq:EHubbard_FF_Action_Fermion}
\end{align}
In addition, we have introduced the free energy ${\mathcal{F}_{c}(J^\varsigma_{\bf{Q}}) = - \ln {({\cal Z}_{c}})}/{\beta N}$, where ${\cal Z}_{c}$ is the partition function of the action $\mathcal{S}_{c}$.
We finally note that writing the initial~\eqref{Eq:EHubbard_Action} and the trial~\eqref{Eq:EHubbard_FF_Action} actions in terms of $\rho^{\varsigma}$ variables above allows us to keep the bare Green's function ${\cal G}_{{\bf k}\nu}$ identical in both actions, simplifying the variational treatment.
In contrast, another choice of variables would necessitate a shift in the chemical potential in the trial action $\mathcal{S}^{*}$ relative the extended Hubbard action $\mathcal{S}$.

For the evaluation of $\langle \ldots \rangle_{{\cal S}_c}$, we explicitly rewrite the expectation value as (see Ref.~\onlinecite{PhysRevB.102.224423} for details):
\begin{equation}
    \langle ... \rangle_{{\cal S}_c} = \langle \langle ... \rangle_{{\cal S}_e}\rangle_{{\cal S}_\phi},\label{Eq:ExpectionValue}
\end{equation}
where the inner expectation value is taken with respect to the fermionic part of the trial action~\eqref{Eq:EHubbard_FF_Action}:
\begin{align}
\mathcal{S}_{e} & =  - \frac{1}{\beta N}\sum_{\bf{k},\nu,\sigma} c_{\bf{k}\nu\sigma}^* \mathcal{G}_{\bf{k}\nu}^{-1} c^{\phantom{*}}_{\bf{k}\nu\sigma} + \sum_{\bf{Q},\varsigma} \phi_{\bf{Q}}^\varsigma \rho^\varsigma_{-\bf{Q}},
\label{Eq:EHubbard_FF_Action_Fermion_Inner}
\end{align}
which depends on the classical fields $\phi_{\bf{Q}}^\varsigma$.
A useful property of the inner expectation value is that Wick's theorem applies, as $\mathcal{S}_{e}$ is a Gaussian action with respect to the fermions. 
Note that for any non-zero value of the classical field $\phi_{\bf{Q}}^\varsigma$ the term $\phi_{\bf{Q}}^\varsigma \rho^\varsigma_{-\bf{Q}}$ in the action~\eqref{Eq:EHubbard_FF_Action_Fermion_Inner} allows for the collective fluctuations by breaking the associate symmetries in the $\mathcal{S}_{e}$ sub-system. 
The symmetries of the full system $\mathcal{S}_{c}$ are, however, retained by ultimately taking the outer expectation value ${\langle \ldots \rangle_{{\cal S}_{\phi}}}$.

In the current work, we limit our considerations to the collective AFM and CDW fluctuations with ${\bf{Q} = (\pi,\pi)}$ wave vector that are the leading mode in the half-filled extended Hubbard model. 
Our choice to keep only the main ${\bf Q}$ mode for each fluctuation on the grounds that the momentum-space representation for the static lattice susceptibility ${X^{\varsigma}({\bf q},\omega=0)}$ at the transition point between the normal and the ordered phases usually has the form of a delta-function-like Bragg peak located at the ordering vectors ${X^{\varsigma}({\bf q},\omega=0)\sim\delta_{\bf q, Q}}$ (see, e.g. Refs.~\cite{stepanov2021coexisting, Vandelli}).
Thus (while a multi-mode FF has been developed to incorporate the leading and sub-leading momentum modes in~\cite{PhysRevB.105.035118}) we argue that considering only the leading $\bf{Q}$-mode is sufficient for predicting phase boundaries in the case of strong competing fluctuations.

Given the symmetries of the considered model, the charge and spin channels are described by two independent stiffness constants $J^{s}_{\bf{Q}}$ and $J^{c}_{\bf{Q}}$ that can be obtained by minimising the corresponding free energy~\eqref{Eq:F_variational} as:
\begin{align}
\frac{\partial \mathcal{F}(J^\varsigma_{\bf{Q}})}{\partial J^\varsigma_{\bf{Q}}} = 0.
\end{align}
This leads to ${J^s_{\bf{Q}} = -U/2}$ for the stiffness constant in the spin channel, in agreement with the result of the previous work~\cite{PhysRevB.102.224423}, and to ${J^c_{\bf{Q}} = U/2 + V_{\bf{Q}}}$ in the charge channel (see Appendix~\ref{AppendixA} for details). 
Importantly, the employed variational approach avoids the hidden Fierz ambiguity in the decoupling of the on-site Coulomb interaction $U$ between the different fluctuating channels~\cite{PhysRevD.68.025020,PhysRevB.70.125111,jaeckel2002understanding}.
In this regard, it is interesting to note that the obtained values of the stiffness constants $J^{\varsigma}_{\bf{Q}}$ correspond to the form of the bare interaction used in the diagrammatic D-TRILEX approach that resolves the Fierz ambiguity problem in a completely different way~\cite{PhysRevB.100.205115, PhysRevB.103.245123, arxiv.2204.06426}.
At this step, the effective action~\eqref{Eq:EHubbard_FF_Action_FluctuatingFields} is fully defined and can be solved numerically exactly, which allows the approach to respect the underlying symmetry of the system and in addition incorporate non-Gaussian fluctuations non-perturbatively, as will be conducted below.

\subsection{Free energy}
In this section we describe the method employed to investigate the interplay between collective CDW and AFM fluctuations in the extended Hubbard model using the developed MCFF method.
The phase diagram of the system can be determined based on the free energy $\mathcal{F}_{\phi}$ of the effective MCFF action~\eqref{Eq:EHubbard_FF_Action_FluctuatingFields}, which allows us to avoid computing the more complex susceptibilities in the instability channels.
In order to find the phase boundary for the CDW phase, we introduce the free energy $\mathcal{F}(\phi^c)$ for the respective classical field $\phi^{c}_{\bf Q}$ by integrating out the spin degrees of freedom $\phi^{s}_{\bf Q}$ numerically exactly:
\begin{align}
{\cal F}(\phi^{c}) = - \frac{1}{\beta N}\ln \int D[\phi^{s}]\,\exp\big\{-{\cal S}_{\phi}[\phi^{c},\phi^{s}]\big\}.
\end{align}
The free energy of the classical vector spin field $\phi^{s}_{\bf Q}$ can be obtained in a similar way by integrating out the $\phi^{c}_{\bf Q}$ field.
This procedure allows us to construct the free energy for a single channel that, however, fully accounts for the effect of collective fluctuations in the other channel that is integrated out.
The introduced free energy has the stability requirement ${J^\varsigma_{\bf{Q}}<0}$ that ensures that $\mathcal{F}(\phi^{\varsigma})$ has a global minimum for each $\phi^\varsigma$.
This requirement limits the regions in which the different collective fluctuations can be incorporated within the MCFF scheme. 
For the considered extended Hubbard model, the stability requirement for the AFM and CDW fluctuations are ${U>0}$ and ${V>U/8}$, respectively.
With the method for constructing the free energy within the MCFF theory, we may now finally generate the $U,V$ phase diagram for the extended Hubbard model.

\section{Results}

\subsection{Phase diagram in the thermodynamic limit}
\begin{figure}[t!]
\includegraphics[width=1.\linewidth]{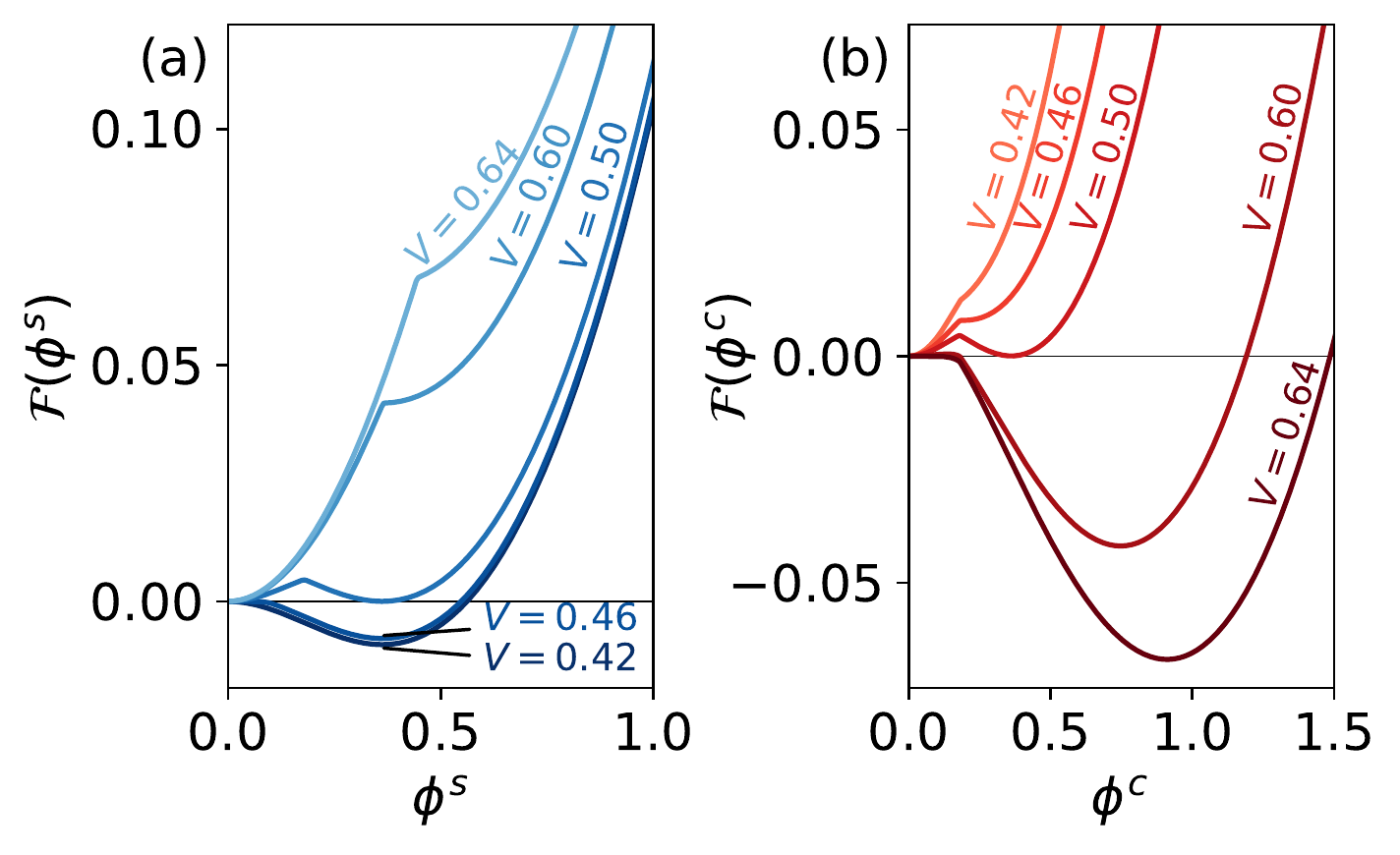}
\caption{Free energy $\mathcal{F}(\phi^\varsigma)$ for the spin (a) and charge (b) channels. 
The results are obtained for the half-filled extended Hubbard model on a square lattice at ${\beta = 10/t}$ and ${U=2t}$ in the vicinity of the CDW-AFM transition point ${V=U/4}$, for a plaquette of $128\times 128$ lattice sites.
Choice of $U,V$ are denoted as stars in Fig.~\ref{figure3}.
\label{figure2}}
\end{figure}

We now focus on the half-filled extended Hubbard model on a square lattice with repulsive $U$ and $V$ interactions. The numerical MCFF investigation is based on the construction of the single-channel free energies in the CDW and AFM channel.
A typical behavior of the introduced single-channel free energy ${\cal F}(\phi^{\varsigma})$ is illustrated in Fig.~\ref{figure2}.
In the normal phase, the global minimum of ${\cal F}(\phi^{\varsigma})$ lies at ${\phi^{\varsigma}=0}$.
The formation of the ordered phase is signaled by a shift of the global minimum to a ${\phi^{\varsigma}\neq0}$ point.
In addition to the global minimum, the free energy may reveal a local minimum that indicates the presence of a metastable phase.
We will discuss the appearance of the metastable phases below.
Finally, we observe a non-analyticity appearing as a kink in the free energy ${\cal F}(\phi^{\varsigma})$. It signals a change of behaviour of ${\cal F}(\phi^{\varsigma})$ between the region in the vicinity of ${\phi^{\varsigma}=0}$, where the fluctuations in the integrated channel are strong, and the region of ${\phi^{\varsigma}\neq0}$, where the fluctuations in the considered channel are strong. Thus, the observed kink is inherently connected to the interplay between the collective CDW and AFM fluctuations.

\begin{figure}[t!]
\includegraphics[width=1.\linewidth]{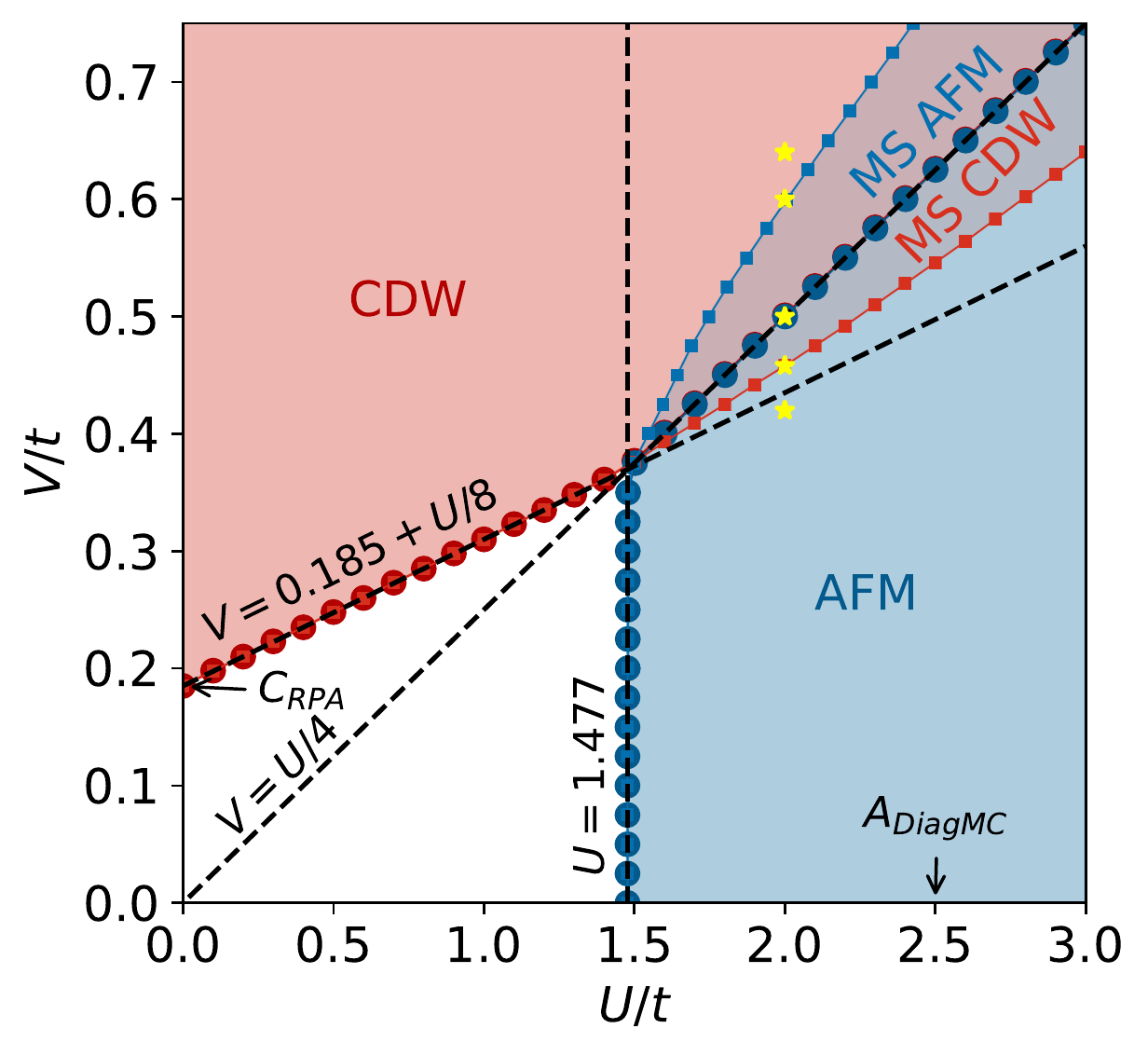}
\caption{Phase diagram for the half-filled extended Hubbard model with repulsive $U,V$ interactions as predicted by the MCFF approach.
The result is obtained at ${\beta=10/t}$ for a plaquette of ${128\times 128}$ lattice sites. 
Red and blue areas depict the CDW and AFM phases, respectively.
The corresponding phase boundaries are shown by colored circles.
Black dashed lines describe the asymptotic behavior of the phase boundaries: ${V=0.185+U/8}$ for CDW, ${U=1.477}$ for AFM, and ${V=U/4}$ for CDW-AFM transitions.
The boundaries of metastable CDW and AFM phases are illustrated by lines with small square markers.
Yellow stars depict the points at which the free energies shown in Fig.~\ref{figure2} were calculated. For comparison, the RPA estimate $C_{\rm RPA}$ for the CDW boundary in the ${U\to0}$, and the DiagMC estimate $A_{\rm DiagMC}$ for the AFM boundary in the limit of ${V\to0}$, taken from Ref.~\cite{PhysRevLett.124.017003}, are included in the thermodynamic limit.
\label{figure3}}
\end{figure}

We perform calculations for a plaquette of ${128\times128}$ lattice sites with periodic boundary conditions, which can be arguably considered as the thermodynamic limit, as we do not see any difference in the results compared to the ${256\times256}$ case.
Fig.~\ref{figure3} shows the phase diagram of the system obtained at ${\beta=10/t}$.
We note that the MCFF method can also be applied at much lower temperatures.
The choice of ${\beta}$ is due to convenience in the comparison to earlier works.
Based on the free energy considerations discussed above, our calculations reveal three phases: a normal (white color), a CDW (red color), and an AFM (blue color) phase.
We find that in the weak coupling regime $U\leq1.447$ the CDW phase boundary follows the ${V = 0.185 + U/8}$ line.
This result is in a perfect agreement with the perturbative estimation ${V = C_{\rm RPA} + U/8}$~\cite{PhysRevB.99.115112}, where the constant $C_{\rm RPA}$ corresponds to the critical value of the non-local interaction for the CDW transition ${V^{\rm CDW}_{U=0}}$ obtained for ${U=0}$ using the random phase approximation (RPA).
The RPA estimate is determined by the critical ${V^{\rm CDW}_{U=0}}$ associate with a singularity in the RPA construction of the charge susceptibility at the $(\pi,\pi)$-point, or equivalently determined by a vanishing RPA dielectric function.
For the considered system, ${C_{\rm RPA}=0.1847}$, which confirms that the MCFF theory correctly captures the exact ${U\to0}$ limit for the CDW phase boundary.
The AFM phase boundary in the weak coupling regime lies along the ${U=1.477}$ line in agreement with the fluctuating local exchange (FLEX) result obtained for ${V=0}$: ${A_{\rm FLEX} = U^{\rm AFM}_{V=0} = 8 C_{\rm RPA}}$. 
However, FLEX is known to underestimate the critical interaction for the AFM transition.
For instance, in the thermodynamic limit the exact diagrammatic Monte Carlo (DiagMC) solution gives ${A_{\rm DiagMC}\simeq2.5}$ for ${\beta=10/t}$~\cite{PhysRevLett.124.017003}.
Determination of $U^{\rm AFM}_{V=0}$ within FLEX is similar to the RPA estimate of the critical ${V^{\rm CDW}_{U=0}}$, associated instead with a divergence of the FLEX construction of the spin susceptibility at the $(\pi,\pi)$-point.
At moderate interaction strengths, if one considers fluctuations only in one channel and completely disregards the other channel, the single-channel FF method predicts the CDW and AFM phase boundary to follow exactly ${V = C_{\rm RPA} + U/8}$ and ${U = 8 C_{\rm RPA}}$, respectively, as depicted by dashed lines in Fig.~\ref{figure3}.
The single-channel FF method thus predicts the weak interaction estimate to continue into the moderate interaction regime. 
If we now consider both fluctuations, the CDW and AFM phases are mutually exclusive, with the interplay leading to the system developing a CDW-AFM phase boundary at ${V=U/4}$ in agreement with the mean-field (RPA or $GW$~\cite{PhysRevB.95.245130}) prediction that was also confirmed by numerically exact techniques~\cite{PhysRevB.39.9397, PhysRevB.42.465}.

\begin{figure}[t!]
\includegraphics[width=1.\linewidth]{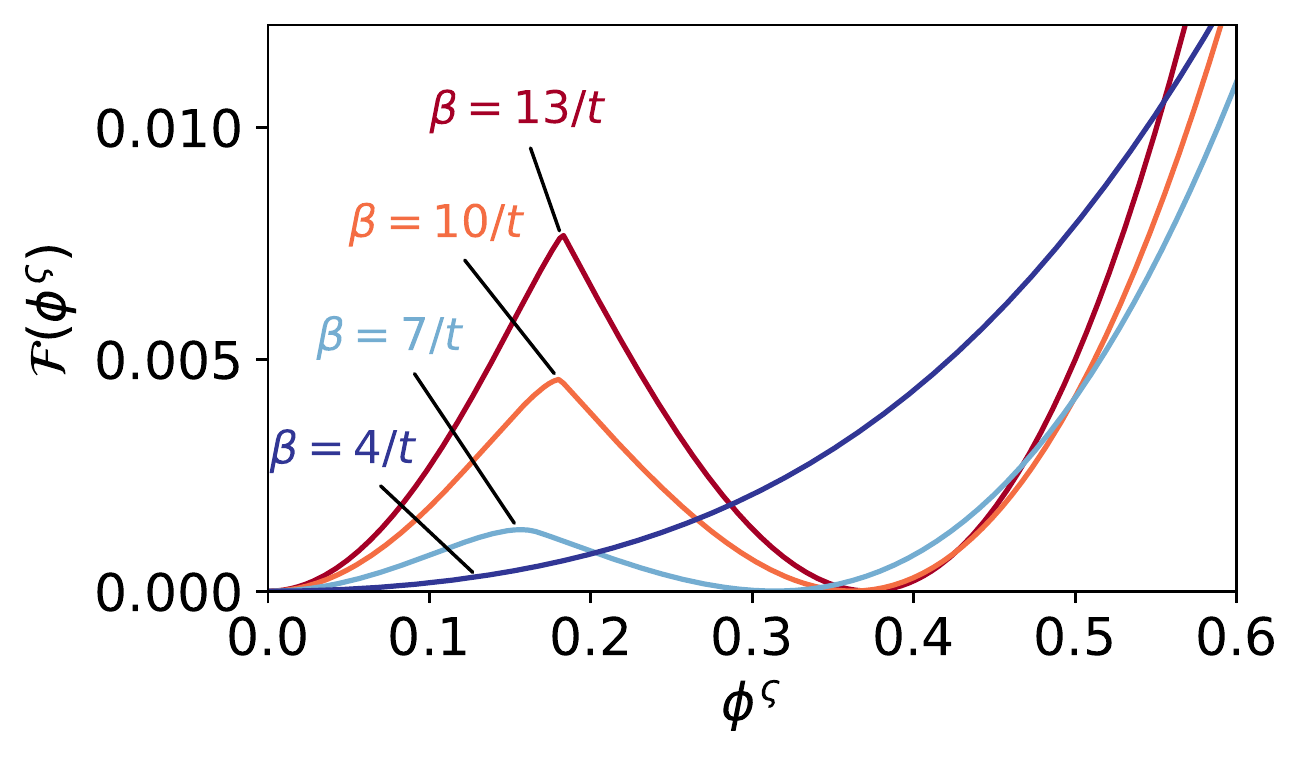}
\caption{Free energy ${\mathcal{F}(\phi^c) = \mathcal{F}(\phi^s)}$ calculated at the CDW-AFM transition point ${U=2t}$, ${V=0.5t}$ for a plaquette of $128\times 128$ lattice sites, with different values of the inverse temperature $\beta$.\label{figure4}}
\end{figure}

Interestingly, we find that in some regions inside the CDW and AFM phases besides the global minimum the free energy ${\cal F}(\phi^{\varsigma})$ reveals a local minimum.
The appearance of the local minimum can be associated with the presence of a metastable (MS) phase.
The boundaries of the metastable phases are depicted in Fig.~\ref{figure3} by red (MS AFM) and blue (MS CDW) lines with small square markers. 
Fig.~\ref{figure2} illustrates a particular example of the free energy behavior in the regime of strong competing CDW and AFM fluctuations.
In the spin channel (Fig.~\ref{figure2}\,a)), as $V$ is increased from deep within the AFM phase the global minimum at ${\phi^s \neq 0}$ in the free energy $\mathcal{F}(\phi^s)$ turns into a local minimum above the CDW-AFM transition point ${V=U/4}$, where the CDW ordering becomes dominant.
The local minimum disappears at the metastable AFM phase transition point, which for ${U=2}$ corresponds to ${V=0.60}$.
Similar results can be found for the charge channel (Fig.~\ref{figure2}\,b)): as $V$ decreases from deep within the CDW phase the metastable CDW phase appears at the AFM-CDW transition point and vanishes at ${U=2}$, ${V=0.46}$.

\begin{figure*}[t!]
\includegraphics[width=0.9\linewidth]{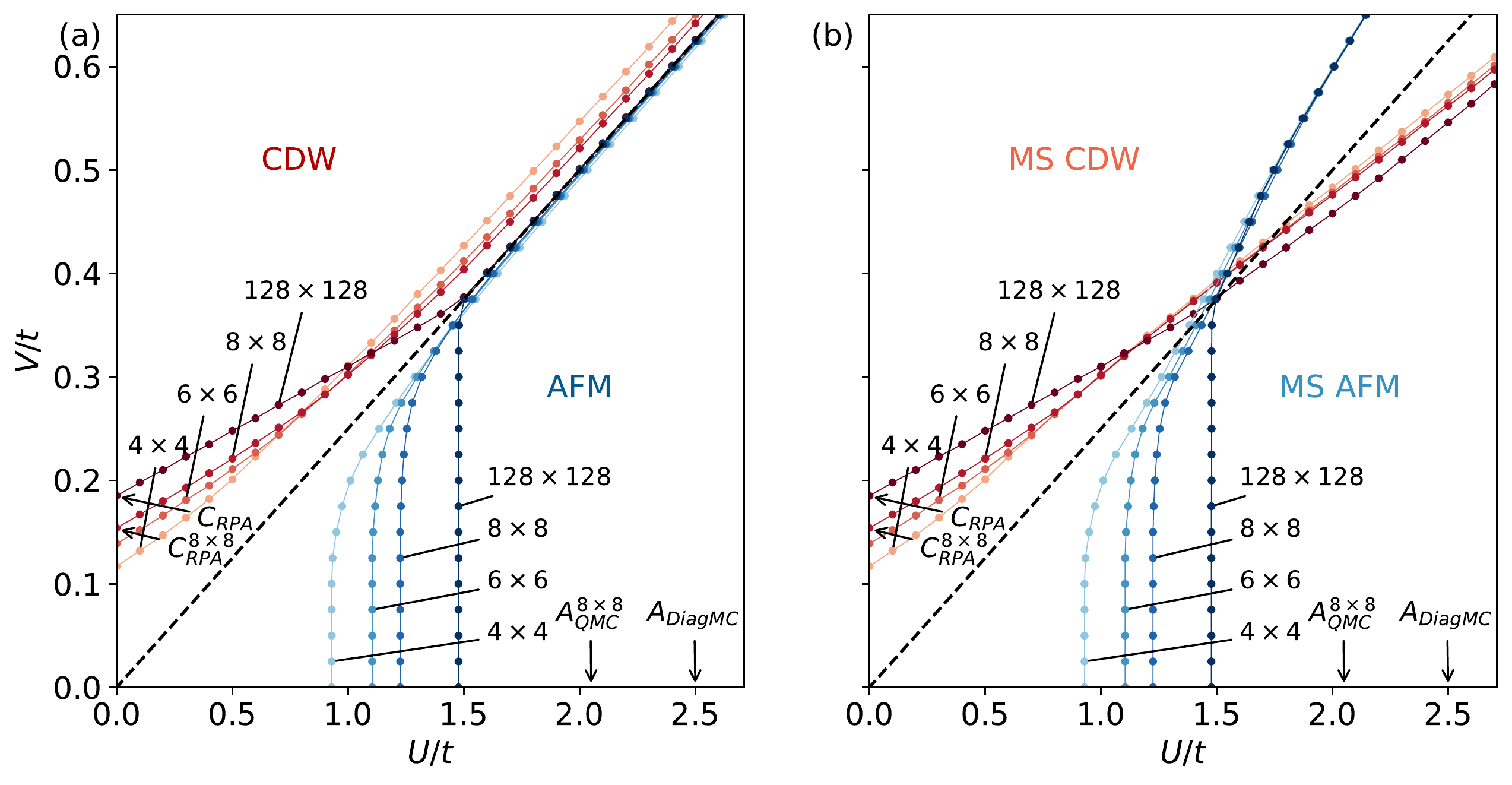}
\caption{Stable (a) and metastable (b) AFM (blue) and CDW (red) ordering boundaries predicted by the MCFF approach for the half-filled extended Hubbard model on ${4\times 4}$, ${6\times 6}$, ${8\times 8}$, and ${128\times 128}$ plaquettes at ${\beta=10}$. 
The dashed line specifies the mean-field estimate for the CDW-AFM phase boundary ${V=U/4}$. 
For comparison, the RPA estimates $C_{\rm RPA}$ and $C^{8 \times 8}_{\rm RPA}$ for the CDW boundary in the ${U\to0}$ in the thermodynamic limit and for a $8\times 8$ plaquette, respectively, are included. 
In addition, the DiagMC estimate $A_{\rm DiagMC}$, taken from Ref.~\cite{PhysRevLett.124.017003}, and the QMC estimate $A_{\rm QMC}^{8\times 8}$ for the AFM boundary in the limit of ${V\to0}$ are included in the thermodynamic limit and for a $8\times 8$ plaquette, respectively, are included.
\label{figure5}}
\end{figure*}

We note that at the CDW-AFM transition the minima located at ${\phi^{\varsigma}=0}$ and ${\phi^{\varsigma}\neq0}$ points correspond to the same value of the free energy $\mathcal{F}(\phi^\varsigma)$.
On the contrary, no metastable solution occurs in the vicinity of the phase boundaries that separate the normal phase from either the CDW or AFM phases.
This result suggests that the transitions in the latter case are of second-order, while the transition between the competing CDW and AFM phases is of first-order. 
In addition, we find that the spin and charge channels are degenerate ($\mathcal{F}(\phi^s)=\mathcal{F}(\phi^c)$) along the CDW-AFM transition line, which indicates that the two instabilities are mutually exclusive.
If these free energies were not identical at the transition point, one channel would be energetically favorable.
Fig.~\ref{figure4} shows the behavior of the free energy $\mathcal{F}(\phi^\varsigma)$ at the CDW-AFM transition point ${U=2}$, ${V=0.5}$ for different temperatures.
We observe that at high temperature corresponding to ${\beta=4/t}$ the AFM and CDW fluctuations are suppressed, and the free energy has only one minimum at ${\phi^{\varsigma}=0}$: the normal phase.
Upon lowering the temperature the second minima develops at ${\phi^{\varsigma} \neq 0}$ and propagates to larger values of $\phi^{\varsigma}$, corresponding to the increase of the strength of corresponding fluctuations. 
We also observe that the free energy barrier between the two minima increases with decreasing temperature.
A larger energy barrier allows for a more stable coexistence of the two phases associated with ${\phi^{\varsigma}=0}$ and ${\phi^{\varsigma}\neq0}$.
It should be emphasized, however, that the two channels are degenerate and that the minima at ${\phi^{\varsigma}=0}$ and ${\phi^{\varsigma}\neq0}$ have the same energy only at the CDW-AFM transition point. 
Away from this point one of the two solutions becomes metastable, which means that one of the CDW or AFM phases always dominates.
Distinguishing between stable and metastable solutions is not a trivial problem, and even the more elaborate DCA method in the regime of strong competing CDW and AFM fluctuations predicts a coexistence between these two mutually exclusive phases~\cite{PhysRevB.99.245146}. Thus, the ability to distinguish between the stable and the metastable phases is an advantage of the MCFF method.

\subsection{Evolution of the phase diagram with the system size}
The MCFF approach can also be applied to small systems, where its performance can be compared to the exact Monte Carlo calculations.
Fig.~\ref{figure5} displays the stable (a) and metastable (b) ordering boundaries for AFM and CDW phases for ${4\times 4}$, ${6\times 6}$ and ${8\times 8}$ plaquettes, in addition to the previously considered ${128\times 128}$ plaquette near the thermodynamic limit.
For all system sizes the MCFF approach extrapolates the AFM and CDW ordering boundaries between weak coupling results obtained respectively on the basis of FLEX calculations and perturbative estimations, and the asymptotic behavior of the CDW-AFM phase boundary at intermediate coupling predicted by mean-field theories. 
A region of coexisting stable and metastable ordering is observed for all system sizes. 
Phase boundaries of the coexistence region appears converged for the ${128\times 128}$ plaquette, indicating its stability in the thermodynamic limit. 

In order to gain insight into the performance of the MCFF approach with inclusion of collective AFM and CO fluctuations, we now perform a comparison with respect to numerically exact QMC simulations.
For a ${8 \times 8}$ plaquette, QMC simulations give us ${A^{8 \times 8}_{\rm QMC} = 2.05 \pm 0.05}$ for ${\beta=10}$. 
By a comparison to the MCFF prediction of ${U^{\rm AFM}_{V=0} = 1.225}$, we find a significant overestimation of the critical interaction $U$ for the AFM phase boundary.
This observation is consistent with our result for the ${128\times 128}$ plaquette and can be related to dynamical correlation effects that are not incorporated within the approach. 
In contrast, the MCFF method accurately determines the CDW phase boundary at small $U$, with the MCFF prediction for the critical interaction ${V^{\rm CDW}_{U=0}}$ coinciding with the RPA result ${C_{\rm RPA}}$ for all plaquettes sizes.

\section{Conclusion}
We have introduced a multi-channel extension of the FF approach to address interplaying collective fluctuations for correlated electronic systems, based on a variational optimization of a trial action respecting the underlying symmetries of the system. 
Exploiting this numerically cheap method, we are able to study competing CDW and AFM fluctuations in the half-filled extended Hubbard model of a large system size. 
The MCFF method predicts a repulsive ${U-V}$ phase diagram in qualitative agreement with more costly methods. 
Our approach correctly captures the ${U\to0}$ limit for the CDW phase boundary, which is a non-trivial problem for computationally heavy cluster-based DMFT techniques due to the cluster size limitations.
In addition, at intermediate interactions a first-order CDW-AFM transition ${U=4V}$ is captured in agreement with numerically exact methods.
A quantitative agreement is observed with respect to DCA simulations~\cite{PhysRevB.99.245146}, with both approaches observing a coexistence region of collective AFM and CDW fluctuations. 
The coexistence regions display a strength of the MCFF approach, as it allows direct access to distinguish between the stable and metastable phases. 
The general nature of the MCFF theory makes it a promising tool for studying the interplay of collective fluctuations in strongly interacting electronic systems.

\begin{acknowledgments}
The authors are thankful to Alexey Rubtsov for inspiring discussions and to Maria Chatzieleftheriou, Benoît Douçot, and Karyn Le Hur for useful comments.
The authors also acknowledge the help of the CPHT computer support team and support from IDRIS/GENCI Orsay under project number A0130901393.
The work of E.A.S was supported by the European Union’s Horizon 2020 Research and Innovation programme under the Marie Sk\l{}odowska Curie grant agreement No.~839551 - \mbox{2DMAGICS}.
\end{acknowledgments}

\appendix

\onecolumngrid

\section{Variational Principle}
\label{AppendixA}

In this appendix we present a detailed derivation of the stiffness constant $J^\varsigma_{\bf{Q}}$. 
To this aim we apply the Peierls-Feynman-Bogoliubov variational principle that maps the initial problem~\eqref{Eq:EHubbard_Action} on the trial action~\eqref{Eq:EHubbard_FF_Action_Fermion} by minimizing the free energy $\mathcal{F}(J^\varsigma_{\bf{Q}})$~\eqref{Eq:F_variational} with respect to variations in $J^\varsigma_{\bf{Q}}$. 
The free energy can be explicitly rewritten as:
\begin{align}
\mathcal{F}(J^\varsigma_{\bf{Q}}) & = \mathcal{F}_{c}(J^\varsigma_{\bf{Q}}) + \frac{1}{\beta N} \left\langle \frac{U}{\beta N}\sum_{{\bf{q}},\omega} \rho_{\bf{q}\omega\uparrow} \rho_{-\bf{q},-\omega\downarrow} + \frac{1}{2}\sum_{{\bf{q}},\omega}\frac{V_{\bf{q}}}{\beta N} \rho_{\bf{q}\omega}\rho_{-\bf{q},-\omega} - \sum_{\varsigma}\frac{1}{2}\frac{J^\varsigma_{\bf{Q}}}{\beta N}\rho_{\bf{Q}}^\varsigma \rho_{\bf{Q}}^\varsigma \right\rangle_{{\cal S}_c}
\label{eq:F_app}
\end{align}
Now exploiting Eq.~\eqref{Eq:ExpectionValue}, we may rewrite the local interaction term explicitly using Wick's theorem as:
\begin{align}
    U \left\langle n_{j\tau\uparrow} n_{j\tau\downarrow}\right\rangle_{{\cal S}_c} & = U \left\langle\left\langle c^\dagger_{j\tau\uparrow}c_{j\tau\uparrow} c^\dagger_{j\tau\downarrow}c_{j\tau\downarrow}\right\rangle_{{\cal S}_e}\right\rangle_{{\cal S}_\phi} \nonumber \\
    & = U \left\langle\left\langle c^\dagger_{j\tau\uparrow}c_{j\tau\uparrow}\right\rangle_{{\cal S}_e} \left\langle c^\dagger_{j\tau\downarrow}c_{j\tau\downarrow}\right\rangle_{{\cal S}_e}-\left\langle c^\dagger_{j\tau\uparrow}c_{j\tau\downarrow}\right\rangle_{{\cal S}_e} \left\langle c^\dagger_{j\tau\downarrow}c_{j\tau\uparrow}\right\rangle_{{\cal S}_e} \right\rangle_{{\cal S}_\phi} \nonumber \\
    & = \frac{U}{4}\left\langle\left\langle n^c_{j\tau}\right\rangle_{{\cal S}_e}^2-\left\langle\vec{n}^s_{j\tau}\right\rangle_{{\cal S}_e}^2 \right\rangle_{{\cal S}_\phi},
\end{align}
where for convenience we have employed a real-space representation for the interaction term.
Rewriting the term in the Fourier basis, we arrive at:
\begin{align}
    \sum_{j,\tau} U \left\langle \rho_{j\tau\uparrow} \rho_{j\tau\downarrow}\right\rangle_{{\cal S}_c} = \frac{1}{\beta N} \sum_{\bf{q},\omega} \frac{U}{4}\left\langle \left\langle \rho^c_{\bf{q}\omega}\right\rangle_{{\cal S}_e} \cdot \left\langle \rho^c_{-\bf{q},-\omega}\right\rangle_{{\cal S}_e} - \left\langle\vec{\rho}^s_{\bf{q}\omega}\right\rangle_{{\cal S}_e} \cdot \left\langle \vec{\rho}^s_{-\bf{q},-\omega}\right\rangle_{{\cal S}_e} \right\rangle_{{\cal S}_\phi}.
\end{align}
Similarly, we may rewrite the non-local interaction term approximately using Wick's theorem as:
\begin{align}
    \frac{1}{2} V_{ij}\left\langle n_{i\tau}n_{j\tau} \right\rangle_{{\cal S}_c} & = \frac{1}{2} V_{ij} \sum_{\sigma\sigma'}\left\langle\left\langle c^\dagger_{i\tau\sigma}c^{\phantom{\dagger}}_{i\tau\sigma}c^\dagger_{j\tau\sigma^\prime}c^{\phantom{\dagger}}_{j\tau\sigma^\prime} \right\rangle_{{\cal S}_e}\right\rangle_{{\cal S}_\phi}\nonumber \\
    & = \frac{1}{2} V_{ij} \sum_{\sigma\sigma'}\left\langle\left\langle c^\dagger_{i\tau\sigma}c_{i\tau\sigma}\right\rangle_{{\cal S}_e}\left\langle c^\dagger_{j\tau\sigma^\prime}c_{j\tau\sigma^\prime} \right\rangle_{{\cal S}_e}- \left\langle c^\dagger_{i\tau\sigma}c_{j\tau\sigma^\prime}\right\rangle_{{\cal S}_e}\left\langle c^\dagger_{j\tau\sigma^\prime}c_{i\tau\sigma} \right\rangle_{{\cal S}_e} \right\rangle_{{\cal S}_\phi}\nonumber \\
    & \approx \frac{1}{2} V_{ij}\left\langle\left\langle n^{c}_{i\tau}\right\rangle_{{\cal S}_e}^2 \right\rangle_{{\cal S}_\phi},
\end{align}
where ${i\neq j}$.
Note that at the last line of this equation we have dropped the sub-leading non-local expectation values scaling as $1/N$, see Ref.~\cite{PhysRevB.102.224423}. Rewriting the term in the Fourier basis, we arrive at:
\begin{align}
    \frac{1}{2}\sum_{ij} V_{ij}\left\langle \rho_{i\tau} \rho_{j\tau} \right\rangle_{{\cal S}_c} & \approx \frac{1}{2\beta N}\sum_{\bf{q},\omega} V_{\bf{q}} \left\langle\left\langle \rho^{c}_{\bf{q}\omega}\right\rangle_{{\cal S}_e} \left\langle \rho^{c}_{-\bf{q},-\omega} \right\rangle_{{\cal S}_e}\right\rangle_{{\cal S}_\phi}.
\end{align}
Similarly, we approximately evaluate the expectation value of the interaction in the MCFF action as:
\begin{align}
    \frac{1}{2}\frac{J_{\bf{Q}}^{\varsigma}}{\beta N}\left\langle {\rho_{\bf{Q}}^\varsigma}^2 \right\rangle_{{\cal S}_c} & \approx \frac{1}{2}\frac{J_{\bf{Q}}^{\varsigma}}{\beta N}\left\langle\left\langle \rho_{\bf{Q}}^\varsigma \right\rangle^2_{{\cal S}_e}\right\rangle_{\cal{S}_\phi}.
\end{align}
The form of the MCFF action ${\cal S}_{e}$~\eqref{Eq:EHubbard_FF_Action_Fermion_Inner} only allows for certain quasi-momentum modes of the local and non-local interaction terms to contribute to the free energy. 
Specifically, only the classical (${\omega=0}$) component with the momentum ${{\bf q = Q}}$ contributes to the average of the shifted density: ${\langle \rho^{\varsigma}_{{\bf q}\omega} \rangle_{{\cal S}_{c}} = \langle \rho^{\varsigma}_{{\bf Q}} \rangle_{{\cal S}_{c}}}$.
Thus, the free energy~\eqref{eq:F_app} takes the following form:
\begin{align}
    \mathcal{F}(J^\varsigma_{\bf{Q}}) & \approx \mathcal{F}_{c}(J^\varsigma_{\bf{Q}}) + \frac{1}{(\beta N)^{2}}\left(\frac{U}{4} + \frac{V_{\bf{Q}}}{2} -\frac{J_{\bf{Q}}^c}{2} \right) \left\langle\left\langle \rho^{c}_{{\bf{Q}}}\right\rangle_{{\cal S}_e}^2\right\rangle_{{\cal S}_\phi} - \frac{1}{(\beta N)^2} \left( \frac{U}{4} + \frac{J_{\bf{Q}}^s}{2} \right) \left\langle\left\langle \vec{\rho}^s_{{\bf{Q}}}\right\rangle_{
    {\cal S}_e}^2 \right\rangle_{{\cal S}_\phi}.
\end{align}
The stiffnesses $J^\varsigma_{\bf{Q}}$ may now be constructed by the proposed variational approach, by varying the free energy $\mathcal{F}(J^\varsigma_{\bf{Q}})$ with respect to $J^\varsigma_{\bf{Q}}$, i.e. ${\partial \mathcal{F}/\partial J^\varsigma_{\bf{Q}} = 0}$. We thus identify ${J^s_{\bf{Q}} = -\frac{U}{2}}$ and ${J^c_{\bf{Q}} = \frac{U}{2} + V_{\bf{Q}}}$. This completes the determination of the stiffness constants $J^\varsigma_{\bf{Q}}$.

\twocolumngrid

\bibliography{Bib_MCFF}

\begin{thebibliography}{77}%
\makeatletter
\providecommand \@ifxundefined [1]{%
 \@ifx{#1\undefined}
}%
\providecommand \@ifnum [1]{%
 \ifnum #1\expandafter \@firstoftwo
 \else \expandafter \@secondoftwo
 \fi
}%
\providecommand \@ifx [1]{%
 \ifx #1\expandafter \@firstoftwo
 \else \expandafter \@secondoftwo
 \fi
}%
\providecommand \natexlab [1]{#1}%
\providecommand \enquote  [1]{``#1''}%
\providecommand \bibnamefont  [1]{#1}%
\providecommand \bibfnamefont [1]{#1}%
\providecommand \citenamefont [1]{#1}%
\providecommand \href@noop [0]{\@secondoftwo}%
\providecommand \href [0]{\begingroup \@sanitize@url \@href}%
\providecommand \@href[1]{\@@startlink{#1}\@@href}%
\providecommand \@@href[1]{\endgroup#1\@@endlink}%
\providecommand \@sanitize@url [0]{\catcode `\\12\catcode `\$12\catcode
  `\&12\catcode `\#12\catcode `\^12\catcode `\_12\catcode `\%12\relax}%
\providecommand \@@startlink[1]{}%
\providecommand \@@endlink[0]{}%
\providecommand \url  [0]{\begingroup\@sanitize@url \@url }%
\providecommand \@url [1]{\endgroup\@href {#1}{\urlprefix }}%
\providecommand \urlprefix  [0]{URL }%
\providecommand \Eprint [0]{\href }%
\providecommand \doibase [0]{http://dx.doi.org/}%
\providecommand \selectlanguage [0]{\@gobble}%
\providecommand \bibinfo  [0]{\@secondoftwo}%
\providecommand \bibfield  [0]{\@secondoftwo}%
\providecommand \translation [1]{[#1]}%
\providecommand \BibitemOpen [0]{}%
\providecommand \bibitemStop [0]{}%
\providecommand \bibitemNoStop [0]{.\EOS\space}%
\providecommand \EOS [0]{\spacefactor3000\relax}%
\providecommand \BibitemShut  [1]{\csname bibitem#1\endcsname}%
\let\auto@bib@innerbib\@empty
\bibitem [{\citenamefont {Lanczos}(1950)}]{jres.045.026}%
  \BibitemOpen
  \bibfield  {author} {\bibinfo {author} {\bibfnamefont {C.}~\bibnamefont
  {Lanczos}},\ }\bibfield  {title} {\enquote {\bibinfo {title} {{An Iteration
  Method for the Solution of the Eigenvalue Problem of Linear Differential and
  Integral Operators}},}\ }\href {\doibase
  https://doi.org/10.6028/jres.045.026} {\bibfield  {journal} {\bibinfo
  {journal} {Journal of Research of the National Bureau of Standards}\ }\textbf
  {\bibinfo {volume} {45}},\ \bibinfo {pages} {255--282} (\bibinfo {year}
  {1950})}\BibitemShut {NoStop}%
\bibitem [{\citenamefont {Hirsch}\ and\ \citenamefont
  {Fye}(1986)}]{PhysRevLett.56.2521}%
  \BibitemOpen
  \bibfield  {author} {\bibinfo {author} {\bibfnamefont {J.~E.}\ \bibnamefont
  {Hirsch}}\ and\ \bibinfo {author} {\bibfnamefont {R.~M.}\ \bibnamefont
  {Fye}},\ }\bibfield  {title} {\enquote {\bibinfo {title} {{Monte Carlo Method
  for Magnetic Impurities in Metals}},}\ }\href {\doibase
  10.1103/PhysRevLett.56.2521} {\bibfield  {journal} {\bibinfo  {journal}
  {Phys. Rev. Lett.}\ }\textbf {\bibinfo {volume} {56}},\ \bibinfo {pages}
  {2521--2524} (\bibinfo {year} {1986})}\BibitemShut {NoStop}%
\bibitem [{\citenamefont {Salmhofer}\ \emph {et~al.}(2004)\citenamefont
  {Salmhofer}, \citenamefont {Honerkamp}, \citenamefont {Metzner},\ and\
  \citenamefont {Lauscher}}]{10.1143/PTP.112.943}%
  \BibitemOpen
  \bibfield  {author} {\bibinfo {author} {\bibfnamefont {M.}~\bibnamefont
  {Salmhofer}}, \bibinfo {author} {\bibfnamefont {C.}~\bibnamefont
  {Honerkamp}}, \bibinfo {author} {\bibfnamefont {W.}~\bibnamefont {Metzner}},
  \ and\ \bibinfo {author} {\bibfnamefont {O.}~\bibnamefont {Lauscher}},\
  }\bibfield  {title} {\enquote {\bibinfo {title} {{Renormalization Group Flows
  into Phases with Broken Symmetry}},}\ }\href {\doibase 10.1143/PTP.112.943}
  {\bibfield  {journal} {\bibinfo  {journal} {Progress of Theoretical Physics}\
  }\textbf {\bibinfo {volume} {112}},\ \bibinfo {pages} {943--970} (\bibinfo
  {year} {2004})}\BibitemShut {NoStop}%
\bibitem [{\citenamefont {Lykos}\ and\ \citenamefont
  {Pratt}(1963)}]{RevModPhys.35.496}%
  \BibitemOpen
  \bibfield  {author} {\bibinfo {author} {\bibfnamefont {P.}~\bibnamefont
  {Lykos}}\ and\ \bibinfo {author} {\bibfnamefont {G.~W.}\ \bibnamefont
  {Pratt}},\ }\bibfield  {title} {\enquote {\bibinfo {title} {{Discussion on
  The Hartree-Fock Approximation}},}\ }\href {\doibase
  10.1103/RevModPhys.35.496} {\bibfield  {journal} {\bibinfo  {journal} {Rev.
  Mod. Phys.}\ }\textbf {\bibinfo {volume} {35}},\ \bibinfo {pages} {496--501}
  (\bibinfo {year} {1963})}\BibitemShut {NoStop}%
\bibitem [{\citenamefont {Hubbard}(1963)}]{rspa.1963.0204}%
  \BibitemOpen
  \bibfield  {author} {\bibinfo {author} {\bibfnamefont {J.}~\bibnamefont
  {Hubbard}},\ }\bibfield  {title} {\enquote {\bibinfo {title} {{Electron
  correlations in narrow energy bands}},}\ }\href {\doibase
  10.1098/rspa.1963.0204} {\bibfield  {journal} {\bibinfo  {journal} {Proc. R.
  Soc. Lond. A}\ }\textbf {\bibinfo {volume} {276}},\ \bibinfo {pages}
  {238–257} (\bibinfo {year} {1963})}\BibitemShut {NoStop}%
\bibitem [{\citenamefont {Gutzwiller}(1963)}]{PhysRevLett.10.159}%
  \BibitemOpen
  \bibfield  {author} {\bibinfo {author} {\bibfnamefont {M.~C.}\ \bibnamefont
  {Gutzwiller}},\ }\bibfield  {title} {\enquote {\bibinfo {title} {{Effect of
  Correlation on the Ferromagnetism of Transition Metals}},}\ }\href {\doibase
  10.1103/PhysRevLett.10.159} {\bibfield  {journal} {\bibinfo  {journal} {Phys.
  Rev. Lett.}\ }\textbf {\bibinfo {volume} {10}},\ \bibinfo {pages} {159--162}
  (\bibinfo {year} {1963})}\BibitemShut {NoStop}%
\bibitem [{\citenamefont {Kanamori}(1963)}]{10.1143/PTP.30.275}%
  \BibitemOpen
  \bibfield  {author} {\bibinfo {author} {\bibfnamefont {J.}~\bibnamefont
  {Kanamori}},\ }\bibfield  {title} {\enquote {\bibinfo {title} {{Electron
  Correlation and Ferromagnetism of Transition Metals}},}\ }\href {\doibase
  10.1143/PTP.30.275} {\bibfield  {journal} {\bibinfo  {journal} {Prog. Theor.
  Phys.}\ }\textbf {\bibinfo {volume} {30}},\ \bibinfo {pages} {275--289}
  (\bibinfo {year} {1963})}\BibitemShut {NoStop}%
\bibitem [{\citenamefont {Hubbard}(1964)}]{rspa.1964.0190}%
  \BibitemOpen
  \bibfield  {author} {\bibinfo {author} {\bibfnamefont {J.}~\bibnamefont
  {Hubbard}},\ }\bibfield  {title} {\enquote {\bibinfo {title} {{Electron
  correlations in narrow energy bands III. An improved solution}},}\ }\href
  {\doibase 10.1098/rspa.1964.0190} {\bibfield  {journal} {\bibinfo  {journal}
  {Proc. R. Soc. Lond. A}\ }\textbf {\bibinfo {volume} {281}},\ \bibinfo
  {pages} {401–419} (\bibinfo {year} {1964})}\BibitemShut {NoStop}%
\bibitem [{\citenamefont {Harris}\ and\ \citenamefont
  {Lange}(1967)}]{PhysRev.157.295}%
  \BibitemOpen
  \bibfield  {author} {\bibinfo {author} {\bibfnamefont {A.~B.}\ \bibnamefont
  {Harris}}\ and\ \bibinfo {author} {\bibfnamefont {R.~V.}\ \bibnamefont
  {Lange}},\ }\bibfield  {title} {\enquote {\bibinfo {title} {{Single-Particle
  Excitations in Narrow Energy Bands}},}\ }\href {\doibase
  10.1103/PhysRev.157.295} {\bibfield  {journal} {\bibinfo  {journal} {Phys.
  Rev.}\ }\textbf {\bibinfo {volume} {157}},\ \bibinfo {pages} {295--314}
  (\bibinfo {year} {1967})}\BibitemShut {NoStop}%
\bibitem [{\citenamefont {Bari}(1971)}]{PhysRevB.3.2662}%
  \BibitemOpen
  \bibfield  {author} {\bibinfo {author} {\bibfnamefont {R.~A.}\ \bibnamefont
  {Bari}},\ }\bibfield  {title} {\enquote {\bibinfo {title} {{Effects of
  Short-Range Interactions on Electron-Charge Ordering and Lattice Distortions
  in the Localized State}},}\ }\href {\doibase 10.1103/PhysRevB.3.2662}
  {\bibfield  {journal} {\bibinfo  {journal} {Phys. Rev. B}\ }\textbf {\bibinfo
  {volume} {3}},\ \bibinfo {pages} {2662--2670} (\bibinfo {year}
  {1971})}\BibitemShut {NoStop}%
\bibitem [{\citenamefont {Vonsovsky}\ and\ \citenamefont
  {Katsnelson}(1979)}]{Vonsovsky_1979}%
  \BibitemOpen
  \bibfield  {author} {\bibinfo {author} {\bibfnamefont {S.~V.}\ \bibnamefont
  {Vonsovsky}}\ and\ \bibinfo {author} {\bibfnamefont {M.~I.}\ \bibnamefont
  {Katsnelson}},\ }\bibfield  {title} {\enquote {\bibinfo {title} {{Some types
  of instabilities in the electron energy spectrum of the polar model of the
  crystal. I. The maximum-polarity state}},}\ }\href {\doibase
  10.1088/0022-3719/12/11/015} {\bibfield  {journal} {\bibinfo  {journal} {J.
  Phys. C: Solid State Phys.}\ }\textbf {\bibinfo {volume} {12}},\ \bibinfo
  {pages} {2043--2053} (\bibinfo {year} {1979})}\BibitemShut {NoStop}%
\bibitem [{\citenamefont {Emery}(1976)}]{PhysRevB.14.2989}%
  \BibitemOpen
  \bibfield  {author} {\bibinfo {author} {\bibfnamefont {V.~J.}\ \bibnamefont
  {Emery}},\ }\bibfield  {title} {\enquote {\bibinfo {title} {{Theory of the
  quasi-one-dimensional electron gas with strong ``on-site'' interactions}},}\
  }\href {\doibase 10.1103/PhysRevB.14.2989} {\bibfield  {journal} {\bibinfo
  {journal} {Phys. Rev. B}\ }\textbf {\bibinfo {volume} {14}},\ \bibinfo
  {pages} {2989--2994} (\bibinfo {year} {1976})}\BibitemShut {NoStop}%
\bibitem [{\citenamefont {S\'olyom}(1979)}]{doi:10.1080/00018737900101375}%
  \BibitemOpen
  \bibfield  {author} {\bibinfo {author} {\bibfnamefont {J.}~\bibnamefont
  {S\'olyom}},\ }\bibfield  {title} {\enquote {\bibinfo {title} {{The Fermi gas
  model of one-dimensional conductors}},}\ }\href {\doibase
  10.1080/00018737900101375} {\bibfield  {journal} {\bibinfo  {journal} {Adv.
  Phys.}\ }\textbf {\bibinfo {volume} {28}},\ \bibinfo {pages} {201--303}
  (\bibinfo {year} {1979})}\BibitemShut {NoStop}%
\bibitem [{\citenamefont {Emery}(1979)}]{emery1979}%
  \BibitemOpen
  \bibfield  {author} {\bibinfo {author} {\bibfnamefont {V.~J.}\ \bibnamefont
  {Emery}},\ }\enquote {\bibinfo {title} {{Theory of the One-Dimensional
  Electron Gas}},}\ in\ \href {\doibase 10.1007/978-1-4613-2895-7_6} {\emph
  {\bibinfo {booktitle} {Highly Conducting One-Dimensional Solids}}},\ \bibinfo
  {editor} {edited by\ \bibinfo {editor} {\bibfnamefont {Jozef~T.}\
  \bibnamefont {Devreese}}, \bibinfo {editor} {\bibfnamefont {Roger~P.}\
  \bibnamefont {Evrard}}, \ and\ \bibinfo {editor} {\bibfnamefont {Victor~E.}\
  \bibnamefont {van Doren}}}\ (\bibinfo  {publisher} {Springer US},\ \bibinfo
  {address} {Boston, MA},\ \bibinfo {year} {1979})\ pp.\ \bibinfo {pages}
  {247--303}\BibitemShut {NoStop}%
\bibitem [{\citenamefont {Fourcade}\ and\ \citenamefont
  {Spronken}(1984{\natexlab{a}})}]{PhysRevB.29.5096}%
  \BibitemOpen
  \bibfield  {author} {\bibinfo {author} {\bibfnamefont {B.}~\bibnamefont
  {Fourcade}}\ and\ \bibinfo {author} {\bibfnamefont {G.}~\bibnamefont
  {Spronken}},\ }\bibfield  {title} {\enquote {\bibinfo {title} {{Real-space
  scaling methods applied to the one-dimensional extended Hubbard model. II.
  The finite-cell scaling method}},}\ }\href {\doibase
  10.1103/PhysRevB.29.5096} {\bibfield  {journal} {\bibinfo  {journal} {Phys.
  Rev. B}\ }\textbf {\bibinfo {volume} {29}},\ \bibinfo {pages} {5096--5102}
  (\bibinfo {year} {1984}{\natexlab{a}})}\BibitemShut {NoStop}%
\bibitem [{\citenamefont {Hirsch}(1984)}]{PhysRevLett.53.2327}%
  \BibitemOpen
  \bibfield  {author} {\bibinfo {author} {\bibfnamefont {J.~E.}\ \bibnamefont
  {Hirsch}},\ }\bibfield  {title} {\enquote {\bibinfo {title}
  {{Charge-Density-Wave to Spin-Density-Wave Transition in the Extended Hubbard
  Model}},}\ }\href {\doibase 10.1103/PhysRevLett.53.2327} {\bibfield
  {journal} {\bibinfo  {journal} {Phys. Rev. Lett.}\ }\textbf {\bibinfo
  {volume} {53}},\ \bibinfo {pages} {2327--2330} (\bibinfo {year}
  {1984})}\BibitemShut {NoStop}%
\bibitem [{\citenamefont {Nakamura}(1999)}]{doi:10.1143/JPSJ.68.3123}%
  \BibitemOpen
  \bibfield  {author} {\bibinfo {author} {\bibfnamefont {M.}~\bibnamefont
  {Nakamura}},\ }\bibfield  {title} {\enquote {\bibinfo {title} {{Mechanism of
  CDW-SDW Transition in One Dimension}},}\ }\href {\doibase
  10.1143/JPSJ.68.3123} {\bibfield  {journal} {\bibinfo  {journal} {J. Phys.
  Soc. Jpn.}\ }\textbf {\bibinfo {volume} {68}},\ \bibinfo {pages} {3123--3126}
  (\bibinfo {year} {1999})}\BibitemShut {NoStop}%
\bibitem [{\citenamefont {Nakamura}(2000)}]{PhysRevB.61.16377}%
  \BibitemOpen
  \bibfield  {author} {\bibinfo {author} {\bibfnamefont {M.}~\bibnamefont
  {Nakamura}},\ }\bibfield  {title} {\enquote {\bibinfo {title} {{Tricritical
  behavior in the extended Hubbard chains}},}\ }\href {\doibase
  10.1103/PhysRevB.61.16377} {\bibfield  {journal} {\bibinfo  {journal} {Phys.
  Rev. B}\ }\textbf {\bibinfo {volume} {61}},\ \bibinfo {pages} {16377--16392}
  (\bibinfo {year} {2000})}\BibitemShut {NoStop}%
\bibitem [{\citenamefont {Fourcade}\ and\ \citenamefont
  {Spronken}(1984{\natexlab{b}})}]{PhysRevB.29.5089}%
  \BibitemOpen
  \bibfield  {author} {\bibinfo {author} {\bibfnamefont {B.}~\bibnamefont
  {Fourcade}}\ and\ \bibinfo {author} {\bibfnamefont {G.}~\bibnamefont
  {Spronken}},\ }\bibfield  {title} {\enquote {\bibinfo {title} {{Real-space
  scaling methods applied to the one-dimensional extended Hubbard model. I. The
  real-space renormalization-group method}},}\ }\href {\doibase
  10.1103/PhysRevB.29.5089} {\bibfield  {journal} {\bibinfo  {journal} {Phys.
  Rev. B}\ }\textbf {\bibinfo {volume} {29}},\ \bibinfo {pages} {5089--5095}
  (\bibinfo {year} {1984}{\natexlab{b}})}\BibitemShut {NoStop}%
\bibitem [{\citenamefont {Zhang}\ and\ \citenamefont
  {Callaway}(1989)}]{PhysRevB.39.9397}%
  \BibitemOpen
  \bibfield  {author} {\bibinfo {author} {\bibfnamefont {Y.}~\bibnamefont
  {Zhang}}\ and\ \bibinfo {author} {\bibfnamefont {J.}~\bibnamefont
  {Callaway}},\ }\bibfield  {title} {\enquote {\bibinfo {title} {{Extended
  Hubbard model in two dimensions}},}\ }\href {\doibase
  10.1103/PhysRevB.39.9397} {\bibfield  {journal} {\bibinfo  {journal} {Phys.
  Rev. B}\ }\textbf {\bibinfo {volume} {39}},\ \bibinfo {pages} {9397--9404}
  (\bibinfo {year} {1989})}\BibitemShut {NoStop}%
\bibitem [{\citenamefont {Callaway}\ \emph {et~al.}(1990)\citenamefont
  {Callaway}, \citenamefont {Chen}, \citenamefont {Kanhere},\ and\
  \citenamefont {Li}}]{PhysRevB.42.465}%
  \BibitemOpen
  \bibfield  {author} {\bibinfo {author} {\bibfnamefont {J.}~\bibnamefont
  {Callaway}}, \bibinfo {author} {\bibfnamefont {D.~P.}\ \bibnamefont {Chen}},
  \bibinfo {author} {\bibfnamefont {D.~G.}\ \bibnamefont {Kanhere}}, \ and\
  \bibinfo {author} {\bibfnamefont {Q.}~\bibnamefont {Li}},\ }\bibfield
  {title} {\enquote {\bibinfo {title} {{Small-cluster calculations for the
  simple and extended Hubbard models}},}\ }\href {\doibase
  10.1103/PhysRevB.42.465} {\bibfield  {journal} {\bibinfo  {journal} {Phys.
  Rev. B}\ }\textbf {\bibinfo {volume} {42}},\ \bibinfo {pages} {465--474}
  (\bibinfo {year} {1990})}\BibitemShut {NoStop}%
\bibitem [{\citenamefont {Yan}(1993)}]{PhysRevB.48.7140}%
  \BibitemOpen
  \bibfield  {author} {\bibinfo {author} {\bibfnamefont {Xin-Zhong}\
  \bibnamefont {Yan}},\ }\bibfield  {title} {\enquote {\bibinfo {title}
  {{Theory of the extended Hubbard model at half filling}},}\ }\href {\doibase
  10.1103/PhysRevB.48.7140} {\bibfield  {journal} {\bibinfo  {journal} {Phys.
  Rev. B}\ }\textbf {\bibinfo {volume} {48}},\ \bibinfo {pages} {7140--7147}
  (\bibinfo {year} {1993})}\BibitemShut {NoStop}%
\bibitem [{\citenamefont {Aichhorn}\ \emph {et~al.}(2004)\citenamefont
  {Aichhorn}, \citenamefont {Evertz}, \citenamefont {von~der Linden},\ and\
  \citenamefont {Potthoff}}]{PhysRevB.70.235107}%
  \BibitemOpen
  \bibfield  {author} {\bibinfo {author} {\bibfnamefont {M.}~\bibnamefont
  {Aichhorn}}, \bibinfo {author} {\bibfnamefont {H.~G.}\ \bibnamefont
  {Evertz}}, \bibinfo {author} {\bibfnamefont {W.}~\bibnamefont {von~der
  Linden}}, \ and\ \bibinfo {author} {\bibfnamefont {M.}~\bibnamefont
  {Potthoff}},\ }\bibfield  {title} {\enquote {\bibinfo {title} {{Charge
  ordering in extended Hubbard models: Variational cluster approach}},}\ }\href
  {\doibase 10.1103/PhysRevB.70.235107} {\bibfield  {journal} {\bibinfo
  {journal} {Phys. Rev. B}\ }\textbf {\bibinfo {volume} {70}},\ \bibinfo
  {pages} {235107} (\bibinfo {year} {2004})}\BibitemShut {NoStop}%
\bibitem [{\citenamefont {Davoudi}\ and\ \citenamefont
  {Tremblay}(2006)}]{PhysRevB.74.035113}%
  \BibitemOpen
  \bibfield  {author} {\bibinfo {author} {\bibfnamefont {B.}~\bibnamefont
  {Davoudi}}\ and\ \bibinfo {author} {\bibfnamefont {A.-M.~S.}\ \bibnamefont
  {Tremblay}},\ }\bibfield  {title} {\enquote {\bibinfo {title}
  {{Nearest-neighbor repulsion and competing charge and spin order in the
  extended Hubbard model}},}\ }\href {\doibase 10.1103/PhysRevB.74.035113}
  {\bibfield  {journal} {\bibinfo  {journal} {Phys. Rev. B}\ }\textbf {\bibinfo
  {volume} {74}},\ \bibinfo {pages} {035113} (\bibinfo {year}
  {2006})}\BibitemShut {NoStop}%
\bibitem [{\citenamefont {Paki}\ \emph {et~al.}(2019)\citenamefont {Paki},
  \citenamefont {Terletska}, \citenamefont {Iskakov},\ and\ \citenamefont
  {Gull}}]{PhysRevB.99.245146}%
  \BibitemOpen
  \bibfield  {author} {\bibinfo {author} {\bibfnamefont {J.}~\bibnamefont
  {Paki}}, \bibinfo {author} {\bibfnamefont {H.}~\bibnamefont {Terletska}},
  \bibinfo {author} {\bibfnamefont {S.}~\bibnamefont {Iskakov}}, \ and\
  \bibinfo {author} {\bibfnamefont {E.}~\bibnamefont {Gull}},\ }\bibfield
  {title} {\enquote {\bibinfo {title} {{Charge order and antiferromagnetism in
  the extended Hubbard model}},}\ }\href {\doibase 10.1103/PhysRevB.99.245146}
  {\bibfield  {journal} {\bibinfo  {journal} {Phys. Rev. B}\ }\textbf {\bibinfo
  {volume} {99}},\ \bibinfo {pages} {245146} (\bibinfo {year}
  {2019})}\BibitemShut {NoStop}%
\bibitem [{\citenamefont {Pudleiner}\ \emph {et~al.}(2019)\citenamefont
  {Pudleiner}, \citenamefont {Kauch}, \citenamefont {Held},\ and\ \citenamefont
  {Li}}]{PhysRevB.100.075108}%
  \BibitemOpen
  \bibfield  {author} {\bibinfo {author} {\bibfnamefont {P.}~\bibnamefont
  {Pudleiner}}, \bibinfo {author} {\bibfnamefont {A.}~\bibnamefont {Kauch}},
  \bibinfo {author} {\bibfnamefont {K.}~\bibnamefont {Held}}, \ and\ \bibinfo
  {author} {\bibfnamefont {G.}~\bibnamefont {Li}},\ }\bibfield  {title}
  {\enquote {\bibinfo {title} {{Competition between antiferromagnetic and
  charge density wave fluctuations in the extended Hubbard model}},}\ }\href
  {\doibase 10.1103/PhysRevB.100.075108} {\bibfield  {journal} {\bibinfo
  {journal} {Phys. Rev. B}\ }\textbf {\bibinfo {volume} {100}},\ \bibinfo
  {pages} {075108} (\bibinfo {year} {2019})}\BibitemShut {NoStop}%
\bibitem [{\citenamefont {Stepanov}\ \emph {et~al.}(2022)\citenamefont
  {Stepanov}, \citenamefont {Harkov}, \citenamefont {R\"osner}, \citenamefont
  {Lichtenstein}, \citenamefont {Katsnelson},\ and\ \citenamefont
  {Rudenko}}]{stepanov2021coexisting}%
  \BibitemOpen
  \bibfield  {author} {\bibinfo {author} {\bibfnamefont {E.~A.}\ \bibnamefont
  {Stepanov}}, \bibinfo {author} {\bibfnamefont {V.}~\bibnamefont {Harkov}},
  \bibinfo {author} {\bibfnamefont {M.}~\bibnamefont {R\"osner}}, \bibinfo
  {author} {\bibfnamefont {A.~I.}\ \bibnamefont {Lichtenstein}}, \bibinfo
  {author} {\bibfnamefont {M.~I.}\ \bibnamefont {Katsnelson}}, \ and\ \bibinfo
  {author} {\bibfnamefont {A.~N.}\ \bibnamefont {Rudenko}},\ }\bibfield
  {title} {\enquote {\bibinfo {title} {{Coexisting charge density wave and
  ferromagnetic instabilities in monolayer InSe}},}\ }\href {\doibase
  10.1038/s41524-022-00798-4} {\bibfield  {journal} {\bibinfo  {journal} {npj
  Comput. Mater.}\ }\textbf {\bibinfo {volume} {8}},\ \bibinfo {pages} {118}
  (\bibinfo {year} {2022})}\BibitemShut {NoStop}%
\bibitem [{\citenamefont {et~al.}(2022)}]{Vandelli}%
  \BibitemOpen
  \bibfield  {author} {\bibinfo {author} {\bibfnamefont {M.~Vandelli}\
  \bibnamefont {et~al.}},\ }\href@noop {} {}\bibinfo {howpublished} {to be
  published} (\bibinfo {year} {2022})\BibitemShut {NoStop}%
\bibitem [{\citenamefont {Ayral}\ \emph {et~al.}(2013)\citenamefont {Ayral},
  \citenamefont {Biermann},\ and\ \citenamefont {Werner}}]{PhysRevB.87.125149}%
  \BibitemOpen
  \bibfield  {author} {\bibinfo {author} {\bibfnamefont {T.}~\bibnamefont
  {Ayral}}, \bibinfo {author} {\bibfnamefont {S.}~\bibnamefont {Biermann}}, \
  and\ \bibinfo {author} {\bibfnamefont {P.}~\bibnamefont {Werner}},\
  }\bibfield  {title} {\enquote {\bibinfo {title} {{Screening and nonlocal
  correlations in the extended Hubbard model from self-consistent combined $GW$
  and dynamical mean field theory}},}\ }\href {\doibase
  10.1103/PhysRevB.87.125149} {\bibfield  {journal} {\bibinfo  {journal} {Phys.
  Rev. B}\ }\textbf {\bibinfo {volume} {87}},\ \bibinfo {pages} {125149}
  (\bibinfo {year} {2013})}\BibitemShut {NoStop}%
\bibitem [{\citenamefont {Hafermann}\ \emph {et~al.}(2014)\citenamefont
  {Hafermann}, \citenamefont {van Loon}, \citenamefont {Katsnelson},
  \citenamefont {Lichtenstein},\ and\ \citenamefont
  {Parcollet}}]{PhysRevB.90.235105}%
  \BibitemOpen
  \bibfield  {author} {\bibinfo {author} {\bibfnamefont {H.}~\bibnamefont
  {Hafermann}}, \bibinfo {author} {\bibfnamefont {E.~G. C.~P.}\ \bibnamefont
  {van Loon}}, \bibinfo {author} {\bibfnamefont {M.~I.}\ \bibnamefont
  {Katsnelson}}, \bibinfo {author} {\bibfnamefont {A.~I.}\ \bibnamefont
  {Lichtenstein}}, \ and\ \bibinfo {author} {\bibfnamefont {O.}~\bibnamefont
  {Parcollet}},\ }\bibfield  {title} {\enquote {\bibinfo {title} {{Collective
  charge excitations of strongly correlated electrons, vertex corrections, and
  gauge invariance}},}\ }\href {\doibase 10.1103/PhysRevB.90.235105} {\bibfield
   {journal} {\bibinfo  {journal} {Phys. Rev. B}\ }\textbf {\bibinfo {volume}
  {90}},\ \bibinfo {pages} {235105} (\bibinfo {year} {2014})}\BibitemShut
  {NoStop}%
\bibitem [{\citenamefont {Terletska}\ \emph {et~al.}(2017)\citenamefont
  {Terletska}, \citenamefont {Chen},\ and\ \citenamefont
  {Gull}}]{PhysRevB.95.115149}%
  \BibitemOpen
  \bibfield  {author} {\bibinfo {author} {\bibfnamefont {H.}~\bibnamefont
  {Terletska}}, \bibinfo {author} {\bibfnamefont {T.}~\bibnamefont {Chen}}, \
  and\ \bibinfo {author} {\bibfnamefont {E.}~\bibnamefont {Gull}},\ }\bibfield
  {title} {\enquote {\bibinfo {title} {{Charge ordering and correlation effects
  in the extended Hubbard model}},}\ }\href {\doibase
  10.1103/PhysRevB.95.115149} {\bibfield  {journal} {\bibinfo  {journal} {Phys.
  Rev. B}\ }\textbf {\bibinfo {volume} {95}},\ \bibinfo {pages} {115149}
  (\bibinfo {year} {2017})}\BibitemShut {NoStop}%
\bibitem [{\citenamefont {van Loon}\ and\ \citenamefont
  {Katsnelson}(2018)}]{van_Loon_2018}%
  \BibitemOpen
  \bibfield  {author} {\bibinfo {author} {\bibfnamefont {E.~G. C.~P.}\
  \bibnamefont {van Loon}}\ and\ \bibinfo {author} {\bibfnamefont {M.~I.}\
  \bibnamefont {Katsnelson}},\ }\bibfield  {title} {\enquote {\bibinfo {title}
  {{The extended Hubbard model with attractive interactions}},}\ }\href
  {\doibase 10.1088/1742-6596/1136/1/012006} {\bibfield  {journal} {\bibinfo
  {journal} {J. Phys.: Conf. Ser.}\ }\textbf {\bibinfo {volume} {1136}},\
  \bibinfo {pages} {012006} (\bibinfo {year} {2018})}\BibitemShut {NoStop}%
\bibitem [{\citenamefont {Katanin}(2019)}]{PhysRevB.99.115112}%
  \BibitemOpen
  \bibfield  {author} {\bibinfo {author} {\bibfnamefont {A.~A.}\ \bibnamefont
  {Katanin}},\ }\bibfield  {title} {\enquote {\bibinfo {title} {{Extended
  dynamical mean field theory combined with the two-particle irreducible
  functional renormalization-group approach as a tool to study strongly
  correlated systems}},}\ }\href {\doibase 10.1103/PhysRevB.99.115112}
  {\bibfield  {journal} {\bibinfo  {journal} {Phys. Rev. B}\ }\textbf {\bibinfo
  {volume} {99}},\ \bibinfo {pages} {115112} (\bibinfo {year}
  {2019})}\BibitemShut {NoStop}%
\bibitem [{\citenamefont {Vandelli}\ \emph {et~al.}(2020)\citenamefont
  {Vandelli}, \citenamefont {Harkov}, \citenamefont {Stepanov}, \citenamefont
  {Gukelberger}, \citenamefont {Kozik}, \citenamefont {Rubio},\ and\
  \citenamefont {Lichtenstein}}]{PhysRevB.102.195109}%
  \BibitemOpen
  \bibfield  {author} {\bibinfo {author} {\bibfnamefont {M.}~\bibnamefont
  {Vandelli}}, \bibinfo {author} {\bibfnamefont {V.}~\bibnamefont {Harkov}},
  \bibinfo {author} {\bibfnamefont {E.~A.}\ \bibnamefont {Stepanov}}, \bibinfo
  {author} {\bibfnamefont {J.}~\bibnamefont {Gukelberger}}, \bibinfo {author}
  {\bibfnamefont {E.}~\bibnamefont {Kozik}}, \bibinfo {author} {\bibfnamefont
  {A.}~\bibnamefont {Rubio}}, \ and\ \bibinfo {author} {\bibfnamefont {A.~I.}\
  \bibnamefont {Lichtenstein}},\ }\bibfield  {title} {\enquote {\bibinfo
  {title} {{Dual boson diagrammatic Monte Carlo approach applied to the
  extended Hubbard model}},}\ }\href {\doibase 10.1103/PhysRevB.102.195109}
  {\bibfield  {journal} {\bibinfo  {journal} {Phys. Rev. B}\ }\textbf {\bibinfo
  {volume} {102}},\ \bibinfo {pages} {195109} (\bibinfo {year}
  {2020})}\BibitemShut {NoStop}%
\bibitem [{\citenamefont {Terletska}\ \emph {et~al.}(2021)\citenamefont
  {Terletska}, \citenamefont {Iskakov}, \citenamefont {Maier},\ and\
  \citenamefont {Gull}}]{PhysRevB.104.085129}%
  \BibitemOpen
  \bibfield  {author} {\bibinfo {author} {\bibfnamefont {H.}~\bibnamefont
  {Terletska}}, \bibinfo {author} {\bibfnamefont {S.}~\bibnamefont {Iskakov}},
  \bibinfo {author} {\bibfnamefont {T.}~\bibnamefont {Maier}}, \ and\ \bibinfo
  {author} {\bibfnamefont {E.}~\bibnamefont {Gull}},\ }\bibfield  {title}
  {\enquote {\bibinfo {title} {{Dynamical cluster approximation study of
  electron localization in the extended Hubbard model}},}\ }\href {\doibase
  10.1103/PhysRevB.104.085129} {\bibfield  {journal} {\bibinfo  {journal}
  {Phys. Rev. B}\ }\textbf {\bibinfo {volume} {104}},\ \bibinfo {pages}
  {085129} (\bibinfo {year} {2021})}\BibitemShut {NoStop}%
\bibitem [{\citenamefont {Stepanov}\ \emph
  {et~al.}(2016{\natexlab{a}})\citenamefont {Stepanov}, \citenamefont {Huber},
  \citenamefont {van Loon}, \citenamefont {Lichtenstein},\ and\ \citenamefont
  {Katsnelson}}]{PhysRevB.94.205110}%
  \BibitemOpen
  \bibfield  {author} {\bibinfo {author} {\bibfnamefont {E.~A.}\ \bibnamefont
  {Stepanov}}, \bibinfo {author} {\bibfnamefont {A.}~\bibnamefont {Huber}},
  \bibinfo {author} {\bibfnamefont {E.~G. C.~P.}\ \bibnamefont {van Loon}},
  \bibinfo {author} {\bibfnamefont {A.~I.}\ \bibnamefont {Lichtenstein}}, \
  and\ \bibinfo {author} {\bibfnamefont {M.~I.}\ \bibnamefont {Katsnelson}},\
  }\bibfield  {title} {\enquote {\bibinfo {title} {{From local to nonlocal
  correlations: The Dual Boson perspective}},}\ }\href {\doibase
  10.1103/PhysRevB.94.205110} {\bibfield  {journal} {\bibinfo  {journal} {Phys.
  Rev. B}\ }\textbf {\bibinfo {volume} {94}},\ \bibinfo {pages} {205110}
  (\bibinfo {year} {2016}{\natexlab{a}})}\BibitemShut {NoStop}%
\bibitem [{\citenamefont {Stepanov}\ \emph
  {et~al.}(2019{\natexlab{a}})\citenamefont {Stepanov}, \citenamefont {Huber},
  \citenamefont {Lichtenstein},\ and\ \citenamefont
  {Katsnelson}}]{PhysRevB.99.115124}%
  \BibitemOpen
  \bibfield  {author} {\bibinfo {author} {\bibfnamefont {E.~A.}\ \bibnamefont
  {Stepanov}}, \bibinfo {author} {\bibfnamefont {A.}~\bibnamefont {Huber}},
  \bibinfo {author} {\bibfnamefont {A.~I.}\ \bibnamefont {Lichtenstein}}, \
  and\ \bibinfo {author} {\bibfnamefont {M.~I.}\ \bibnamefont {Katsnelson}},\
  }\bibfield  {title} {\enquote {\bibinfo {title} {{Effective Ising model for
  correlated systems with charge ordering}},}\ }\href {\doibase
  10.1103/PhysRevB.99.115124} {\bibfield  {journal} {\bibinfo  {journal} {Phys.
  Rev. B}\ }\textbf {\bibinfo {volume} {99}},\ \bibinfo {pages} {115124}
  (\bibinfo {year} {2019}{\natexlab{a}})}\BibitemShut {NoStop}%
\bibitem [{\citenamefont {Hettler}\ \emph {et~al.}(1998)\citenamefont
  {Hettler}, \citenamefont {Tahvildar-Zadeh}, \citenamefont {Jarrell},
  \citenamefont {Pruschke},\ and\ \citenamefont
  {Krishnamurthy}}]{PhysRevB.58.R7475}%
  \BibitemOpen
  \bibfield  {author} {\bibinfo {author} {\bibfnamefont {M.~H.}\ \bibnamefont
  {Hettler}}, \bibinfo {author} {\bibfnamefont {A.~N.}\ \bibnamefont
  {Tahvildar-Zadeh}}, \bibinfo {author} {\bibfnamefont {M.}~\bibnamefont
  {Jarrell}}, \bibinfo {author} {\bibfnamefont {T.}~\bibnamefont {Pruschke}}, \
  and\ \bibinfo {author} {\bibfnamefont {H.~R.}\ \bibnamefont
  {Krishnamurthy}},\ }\bibfield  {title} {\enquote {\bibinfo {title} {{Nonlocal
  dynamical correlations of strongly interacting electron systems}},}\ }\href
  {\doibase 10.1103/PhysRevB.58.R7475} {\bibfield  {journal} {\bibinfo
  {journal} {Phys. Rev. B}\ }\textbf {\bibinfo {volume} {58}},\ \bibinfo
  {pages} {R7475--R7479} (\bibinfo {year} {1998})}\BibitemShut {NoStop}%
\bibitem [{\citenamefont {Hettler}\ \emph {et~al.}(2000)\citenamefont
  {Hettler}, \citenamefont {Mukherjee}, \citenamefont {Jarrell},\ and\
  \citenamefont {Krishnamurthy}}]{PhysRevB.61.12739}%
  \BibitemOpen
  \bibfield  {author} {\bibinfo {author} {\bibfnamefont {M.~H.}\ \bibnamefont
  {Hettler}}, \bibinfo {author} {\bibfnamefont {M.}~\bibnamefont {Mukherjee}},
  \bibinfo {author} {\bibfnamefont {M.}~\bibnamefont {Jarrell}}, \ and\
  \bibinfo {author} {\bibfnamefont {H.~R.}\ \bibnamefont {Krishnamurthy}},\
  }\bibfield  {title} {\enquote {\bibinfo {title} {{Dynamical cluster
  approximation: Nonlocal dynamics of correlated electron systems}},}\ }\href
  {\doibase 10.1103/PhysRevB.61.12739} {\bibfield  {journal} {\bibinfo
  {journal} {Phys. Rev. B}\ }\textbf {\bibinfo {volume} {61}},\ \bibinfo
  {pages} {12739--12756} (\bibinfo {year} {2000})}\BibitemShut {NoStop}%
\bibitem [{\citenamefont {Aryanpour}\ \emph {et~al.}(2002)\citenamefont
  {Aryanpour}, \citenamefont {Hettler},\ and\ \citenamefont
  {Jarrell}}]{PhysRevB.65.153102}%
  \BibitemOpen
  \bibfield  {author} {\bibinfo {author} {\bibfnamefont {K.}~\bibnamefont
  {Aryanpour}}, \bibinfo {author} {\bibfnamefont {M.~H.}\ \bibnamefont
  {Hettler}}, \ and\ \bibinfo {author} {\bibfnamefont {M.}~\bibnamefont
  {Jarrell}},\ }\bibfield  {title} {\enquote {\bibinfo {title} {{Analysis of
  the dynamical cluster approximation for the Hubbard model}},}\ }\href
  {\doibase 10.1103/PhysRevB.65.153102} {\bibfield  {journal} {\bibinfo
  {journal} {Phys. Rev. B}\ }\textbf {\bibinfo {volume} {65}},\ \bibinfo
  {pages} {153102} (\bibinfo {year} {2002})}\BibitemShut {NoStop}%
\bibitem [{\citenamefont {Hohenberg}(1967)}]{PhysRev.158.383}%
  \BibitemOpen
  \bibfield  {author} {\bibinfo {author} {\bibfnamefont {P.~C.}\ \bibnamefont
  {Hohenberg}},\ }\bibfield  {title} {\enquote {\bibinfo {title} {{Existence of
  Long-Range Order in One and Two Dimensions}},}\ }\href {\doibase
  10.1103/PhysRev.158.383} {\bibfield  {journal} {\bibinfo  {journal} {Phys.
  Rev.}\ }\textbf {\bibinfo {volume} {158}},\ \bibinfo {pages} {383--386}
  (\bibinfo {year} {1967})}\BibitemShut {NoStop}%
\bibitem [{\citenamefont {Mermin}\ and\ \citenamefont
  {Wagner}(1966)}]{PhysRevLett.17.1133}%
  \BibitemOpen
  \bibfield  {author} {\bibinfo {author} {\bibfnamefont {N.~D.}\ \bibnamefont
  {Mermin}}\ and\ \bibinfo {author} {\bibfnamefont {H.}~\bibnamefont
  {Wagner}},\ }\bibfield  {title} {\enquote {\bibinfo {title} {{Absence of
  Ferromagnetism or Antiferromagnetism in One- or Two-Dimensional Isotropic
  Heisenberg Models}},}\ }\href {\doibase 10.1103/PhysRevLett.17.1133}
  {\bibfield  {journal} {\bibinfo  {journal} {Phys. Rev. Lett.}\ }\textbf
  {\bibinfo {volume} {17}},\ \bibinfo {pages} {1133--1136} (\bibinfo {year}
  {1966})}\BibitemShut {NoStop}%
\bibitem [{\citenamefont {Walker}\ and\ \citenamefont
  {Ruijgrok}(1968)}]{PhysRev.171.513}%
  \BibitemOpen
  \bibfield  {author} {\bibinfo {author} {\bibfnamefont {M.~B.}\ \bibnamefont
  {Walker}}\ and\ \bibinfo {author} {\bibfnamefont {Th.~W.}\ \bibnamefont
  {Ruijgrok}},\ }\bibfield  {title} {\enquote {\bibinfo {title} {{Absence of
  Magnetic Ordering in One and Two Dimensions in a Many-Band Model for
  Interacting Electrons in a Metal}},}\ }\href {\doibase
  10.1103/PhysRev.171.513} {\bibfield  {journal} {\bibinfo  {journal} {Phys.
  Rev.}\ }\textbf {\bibinfo {volume} {171}},\ \bibinfo {pages} {513--515}
  (\bibinfo {year} {1968})}\BibitemShut {NoStop}%
\bibitem [{\citenamefont {Lichtenstein}\ and\ \citenamefont
  {Katsnelson}(2000)}]{PhysRevB.62.R9283}%
  \BibitemOpen
  \bibfield  {author} {\bibinfo {author} {\bibfnamefont {A.~I.}\ \bibnamefont
  {Lichtenstein}}\ and\ \bibinfo {author} {\bibfnamefont {M.~I.}\ \bibnamefont
  {Katsnelson}},\ }\bibfield  {title} {\enquote {\bibinfo {title}
  {{Antiferromagnetism and d-wave superconductivity in cuprates: A cluster
  dynamical mean-field theory}},}\ }\href {\doibase 10.1103/PhysRevB.62.R9283}
  {\bibfield  {journal} {\bibinfo  {journal} {Phys. Rev. B}\ }\textbf {\bibinfo
  {volume} {62}},\ \bibinfo {pages} {R9283--R9286} (\bibinfo {year}
  {2000})}\BibitemShut {NoStop}%
\bibitem [{\citenamefont {Kotliar}\ \emph {et~al.}(2001)\citenamefont
  {Kotliar}, \citenamefont {Savrasov}, \citenamefont {P\'alsson},\ and\
  \citenamefont {Biroli}}]{PhysRevLett.87.186401}%
  \BibitemOpen
  \bibfield  {author} {\bibinfo {author} {\bibfnamefont {G.}~\bibnamefont
  {Kotliar}}, \bibinfo {author} {\bibfnamefont {S.~Y.}\ \bibnamefont
  {Savrasov}}, \bibinfo {author} {\bibfnamefont {G.}~\bibnamefont {P\'alsson}},
  \ and\ \bibinfo {author} {\bibfnamefont {G.}~\bibnamefont {Biroli}},\
  }\bibfield  {title} {\enquote {\bibinfo {title} {{Cellular Dynamical Mean
  Field Approach to Strongly Correlated Systems}},}\ }\href {\doibase
  10.1103/PhysRevLett.87.186401} {\bibfield  {journal} {\bibinfo  {journal}
  {Phys. Rev. Lett.}\ }\textbf {\bibinfo {volume} {87}},\ \bibinfo {pages}
  {186401} (\bibinfo {year} {2001})}\BibitemShut {NoStop}%
\bibitem [{\citenamefont {Maier}\ \emph {et~al.}(2005)\citenamefont {Maier},
  \citenamefont {Jarrell}, \citenamefont {Pruschke},\ and\ \citenamefont
  {Hettler}}]{RevModPhys.77.1027}%
  \BibitemOpen
  \bibfield  {author} {\bibinfo {author} {\bibfnamefont {T.}~\bibnamefont
  {Maier}}, \bibinfo {author} {\bibfnamefont {M.}~\bibnamefont {Jarrell}},
  \bibinfo {author} {\bibfnamefont {T.}~\bibnamefont {Pruschke}}, \ and\
  \bibinfo {author} {\bibfnamefont {M.~H.}\ \bibnamefont {Hettler}},\
  }\bibfield  {title} {\enquote {\bibinfo {title} {{Quantum cluster
  theories}},}\ }\href {\doibase 10.1103/RevModPhys.77.1027} {\bibfield
  {journal} {\bibinfo  {journal} {Rev. Mod. Phys.}\ }\textbf {\bibinfo {volume}
  {77}},\ \bibinfo {pages} {1027--1080} (\bibinfo {year} {2005})}\BibitemShut
  {NoStop}%
\bibitem [{\citenamefont {Tremblay}\ \emph {et~al.}(2006)\citenamefont
  {Tremblay}, \citenamefont {Kyung},\ and\ \citenamefont
  {S\'en\'echal}}]{doi:10.1063/1.2199446}%
  \BibitemOpen
  \bibfield  {author} {\bibinfo {author} {\bibfnamefont {A.-M.~S.}\
  \bibnamefont {Tremblay}}, \bibinfo {author} {\bibfnamefont {B.}~\bibnamefont
  {Kyung}}, \ and\ \bibinfo {author} {\bibfnamefont {D.}~\bibnamefont
  {S\'en\'echal}},\ }\bibfield  {title} {\enquote {\bibinfo {title} {{Pseudogap
  and high-temperature superconductivity from weak to strong coupling. Towards
  a quantitative theory (Review Article)}},}\ }\href {\doibase
  10.1063/1.2199446} {\bibfield  {journal} {\bibinfo  {journal} {Low Temp.
  Phys.}\ }\textbf {\bibinfo {volume} {32}},\ \bibinfo {pages} {424--451}
  (\bibinfo {year} {2006})}\BibitemShut {NoStop}%
\bibitem [{\citenamefont {Kotliar}\ \emph {et~al.}(2006)\citenamefont
  {Kotliar}, \citenamefont {Savrasov}, \citenamefont {Haule}, \citenamefont
  {Oudovenko}, \citenamefont {Parcollet},\ and\ \citenamefont
  {Marianetti}}]{RevModPhys.78.865}%
  \BibitemOpen
  \bibfield  {author} {\bibinfo {author} {\bibfnamefont {G.}~\bibnamefont
  {Kotliar}}, \bibinfo {author} {\bibfnamefont {S.~Y.}\ \bibnamefont
  {Savrasov}}, \bibinfo {author} {\bibfnamefont {K.}~\bibnamefont {Haule}},
  \bibinfo {author} {\bibfnamefont {V.~S.}\ \bibnamefont {Oudovenko}}, \bibinfo
  {author} {\bibfnamefont {O.}~\bibnamefont {Parcollet}}, \ and\ \bibinfo
  {author} {\bibfnamefont {C.~A.}\ \bibnamefont {Marianetti}},\ }\bibfield
  {title} {\enquote {\bibinfo {title} {{Electronic structure calculations with
  dynamical mean-field theory}},}\ }\href {\doibase 10.1103/RevModPhys.78.865}
  {\bibfield  {journal} {\bibinfo  {journal} {Rev. Mod. Phys.}\ }\textbf
  {\bibinfo {volume} {78}},\ \bibinfo {pages} {865--951} (\bibinfo {year}
  {2006})}\BibitemShut {NoStop}%
\bibitem [{\citenamefont {Harland}\ \emph {et~al.}(2016)\citenamefont
  {Harland}, \citenamefont {Katsnelson},\ and\ \citenamefont
  {Lichtenstein}}]{PhysRevB.94.125133}%
  \BibitemOpen
  \bibfield  {author} {\bibinfo {author} {\bibfnamefont {M.}~\bibnamefont
  {Harland}}, \bibinfo {author} {\bibfnamefont {M.~I.}\ \bibnamefont
  {Katsnelson}}, \ and\ \bibinfo {author} {\bibfnamefont {A.~I.}\ \bibnamefont
  {Lichtenstein}},\ }\bibfield  {title} {\enquote {\bibinfo {title} {{Plaquette
  valence bond theory of high-temperature superconductivity}},}\ }\href
  {\doibase 10.1103/PhysRevB.94.125133} {\bibfield  {journal} {\bibinfo
  {journal} {Phys. Rev. B}\ }\textbf {\bibinfo {volume} {94}},\ \bibinfo
  {pages} {125133} (\bibinfo {year} {2016})}\BibitemShut {NoStop}%
\bibitem [{\citenamefont {Diatlov}\ \emph {et~al.}(1957)\citenamefont
  {Diatlov}, \citenamefont {Sudakov},\ and\ \citenamefont
  {Ter-Martirosian}}]{osti_4338008}%
  \BibitemOpen
  \bibfield  {author} {\bibinfo {author} {\bibfnamefont {I.~T.}\ \bibnamefont
  {Diatlov}}, \bibinfo {author} {\bibfnamefont {V.~V.}\ \bibnamefont
  {Sudakov}}, \ and\ \bibinfo {author} {\bibfnamefont {K.~A.}\ \bibnamefont
  {Ter-Martirosian}},\ }\bibfield  {title} {\enquote {\bibinfo {title}
  {{Asymptotic meson-meson scattering theory}},}\ }\href@noop {} {\bibfield
  {journal} {\bibinfo  {journal} {Soviet Phys. JETP}\ }\textbf {\bibinfo
  {volume} {5}} (\bibinfo {year} {1957})}\BibitemShut {NoStop}%
\bibitem [{\citenamefont {De~Dominicis}(1962)}]{1.1724313}%
  \BibitemOpen
  \bibfield  {author} {\bibinfo {author} {\bibfnamefont {C.}~\bibnamefont
  {De~Dominicis}},\ }\bibfield  {title} {\enquote {\bibinfo {title}
  {{Variational formulations of equilibrium statistical mechanics}},}\ }\href
  {\doibase 10.1063/1.172431} {\bibfield  {journal} {\bibinfo  {journal} {J.
  Math. Phys.}\ }\textbf {\bibinfo {volume} {3}} (\bibinfo {year} {1962}),\
  10.1063/1.172431}\BibitemShut {NoStop}%
\bibitem [{\citenamefont {De~Dominicis}(1964)}]{1.1704062}%
  \BibitemOpen
  \bibfield  {author} {\bibinfo {author} {\bibfnamefont {C.}~\bibnamefont
  {De~Dominicis}},\ }\bibfield  {title} {\enquote {\bibinfo {title}
  {{Stationary entropy principle and renormalization in normal and superfluid
  systems. I. algebraic formulation}},}\ }\href {\doibase 10.1063/1.1704062}
  {\bibfield  {journal} {\bibinfo  {journal} {J. Math. Phys.}\ }\textbf
  {\bibinfo {volume} {5}} (\bibinfo {year} {1964}),\
  10.1063/1.1704062}\BibitemShut {NoStop}%
\bibitem [{\citenamefont {Bickers}\ and\ \citenamefont
  {Scalapino}(1989)}]{BICKERS1989206}%
  \BibitemOpen
  \bibfield  {author} {\bibinfo {author} {\bibfnamefont {N.~E.}\ \bibnamefont
  {Bickers}}\ and\ \bibinfo {author} {\bibfnamefont {D.~J.}\ \bibnamefont
  {Scalapino}},\ }\bibfield  {title} {\enquote {\bibinfo {title} {{Conserving
  approximations for strongly fluctuating electron systems. I. Formalism and
  calculational approach}},}\ }\href {\doibase
  https://doi.org/10.1016/0003-4916(89)90359-X} {\bibfield  {journal} {\bibinfo
   {journal} {Ann. Phys.}\ }\textbf {\bibinfo {volume} {193}},\ \bibinfo
  {pages} {206--251} (\bibinfo {year} {1989})}\BibitemShut {NoStop}%
\bibitem [{\citenamefont {Bickers}\ and\ \citenamefont
  {White}(1991)}]{PhysRevB.43.8044}%
  \BibitemOpen
  \bibfield  {author} {\bibinfo {author} {\bibfnamefont {N.~E.}\ \bibnamefont
  {Bickers}}\ and\ \bibinfo {author} {\bibfnamefont {S.~R.}\ \bibnamefont
  {White}},\ }\bibfield  {title} {\enquote {\bibinfo {title} {{Conserving
  approximations for strongly fluctuating electron systems. II. Numerical
  results and parquet extension}},}\ }\href {\doibase 10.1103/PhysRevB.43.8044}
  {\bibfield  {journal} {\bibinfo  {journal} {Phys. Rev. B}\ }\textbf {\bibinfo
  {volume} {43}},\ \bibinfo {pages} {8044--8064} (\bibinfo {year}
  {1991})}\BibitemShut {NoStop}%
\bibitem [{\citenamefont {Bickers}(2004)}]{Bickers2004}%
  \BibitemOpen
  \bibfield  {author} {\bibinfo {author} {\bibfnamefont {N.~E.}\ \bibnamefont
  {Bickers}},\ }\enquote {\bibinfo {title} {{Self-Consistent Many-Body Theory
  for Condensed Matter Systems}},}\ in\ \href {\doibase
  10.1007/0-387-21717-7_6} {\emph {\bibinfo {booktitle} {{Theoretical Methods
  for Strongly Correlated Electrons}}}}\ (\bibinfo  {publisher} {Springer},\
  \bibinfo {address} {New York},\ \bibinfo {year} {2004})\ pp.\ \bibinfo
  {pages} {237--296}\BibitemShut {NoStop}%
\bibitem [{\citenamefont {Rohringer}\ \emph {et~al.}(2018)\citenamefont
  {Rohringer}, \citenamefont {Hafermann}, \citenamefont {Toschi}, \citenamefont
  {Katanin}, \citenamefont {Antipov}, \citenamefont {Katsnelson}, \citenamefont
  {Lichtenstein}, \citenamefont {Rubtsov},\ and\ \citenamefont
  {Held}}]{RevModPhys.90.025003}%
  \BibitemOpen
  \bibfield  {author} {\bibinfo {author} {\bibfnamefont {G.}~\bibnamefont
  {Rohringer}}, \bibinfo {author} {\bibfnamefont {H.}~\bibnamefont
  {Hafermann}}, \bibinfo {author} {\bibfnamefont {A.}~\bibnamefont {Toschi}},
  \bibinfo {author} {\bibfnamefont {A.~A.}\ \bibnamefont {Katanin}}, \bibinfo
  {author} {\bibfnamefont {A.~E.}\ \bibnamefont {Antipov}}, \bibinfo {author}
  {\bibfnamefont {M.~I.}\ \bibnamefont {Katsnelson}}, \bibinfo {author}
  {\bibfnamefont {A.~I.}\ \bibnamefont {Lichtenstein}}, \bibinfo {author}
  {\bibfnamefont {A.~N.}\ \bibnamefont {Rubtsov}}, \ and\ \bibinfo {author}
  {\bibfnamefont {K.}~\bibnamefont {Held}},\ }\bibfield  {title} {\enquote
  {\bibinfo {title} {{Diagrammatic routes to nonlocal correlations beyond
  dynamical mean field theory}},}\ }\href {\doibase
  10.1103/RevModPhys.90.025003} {\bibfield  {journal} {\bibinfo  {journal}
  {Rev. Mod. Phys.}\ }\textbf {\bibinfo {volume} {90}},\ \bibinfo {pages}
  {025003} (\bibinfo {year} {2018})}\BibitemShut {NoStop}%
\bibitem [{\citenamefont {van Loon}\ \emph {et~al.}(2014)\citenamefont {van
  Loon}, \citenamefont {Lichtenstein}, \citenamefont {Katsnelson},
  \citenamefont {Parcollet},\ and\ \citenamefont
  {Hafermann}}]{PhysRevB.90.235135}%
  \BibitemOpen
  \bibfield  {author} {\bibinfo {author} {\bibfnamefont {E.~G. C.~P.}\
  \bibnamefont {van Loon}}, \bibinfo {author} {\bibfnamefont {A.~I.}\
  \bibnamefont {Lichtenstein}}, \bibinfo {author} {\bibfnamefont {M.~I.}\
  \bibnamefont {Katsnelson}}, \bibinfo {author} {\bibfnamefont
  {O.}~\bibnamefont {Parcollet}}, \ and\ \bibinfo {author} {\bibfnamefont
  {H.}~\bibnamefont {Hafermann}},\ }\bibfield  {title} {\enquote {\bibinfo
  {title} {{Beyond extended dynamical mean-field theory: Dual boson approach to
  the two-dimensional extended Hubbard model}},}\ }\href {\doibase
  10.1103/PhysRevB.90.235135} {\bibfield  {journal} {\bibinfo  {journal} {Phys.
  Rev. B}\ }\textbf {\bibinfo {volume} {90}},\ \bibinfo {pages} {235135}
  (\bibinfo {year} {2014})}\BibitemShut {NoStop}%
\bibitem [{\citenamefont {Stepanov}\ \emph
  {et~al.}(2016{\natexlab{b}})\citenamefont {Stepanov}, \citenamefont {van
  Loon}, \citenamefont {Katanin}, \citenamefont {Lichtenstein}, \citenamefont
  {Katsnelson},\ and\ \citenamefont {Rubtsov}}]{PhysRevB.93.045107}%
  \BibitemOpen
  \bibfield  {author} {\bibinfo {author} {\bibfnamefont {E.~A.}\ \bibnamefont
  {Stepanov}}, \bibinfo {author} {\bibfnamefont {E.~G. C.~P.}\ \bibnamefont
  {van Loon}}, \bibinfo {author} {\bibfnamefont {A.~A.}\ \bibnamefont
  {Katanin}}, \bibinfo {author} {\bibfnamefont {A.~I.}\ \bibnamefont
  {Lichtenstein}}, \bibinfo {author} {\bibfnamefont {M.~I.}\ \bibnamefont
  {Katsnelson}}, \ and\ \bibinfo {author} {\bibfnamefont {A.~N.}\ \bibnamefont
  {Rubtsov}},\ }\bibfield  {title} {\enquote {\bibinfo {title}
  {{Self-consistent dual boson approach to single-particle and collective
  excitations in correlated systems}},}\ }\href {\doibase
  10.1103/PhysRevB.93.045107} {\bibfield  {journal} {\bibinfo  {journal} {Phys.
  Rev. B}\ }\textbf {\bibinfo {volume} {93}},\ \bibinfo {pages} {045107}
  (\bibinfo {year} {2016}{\natexlab{b}})}\BibitemShut {NoStop}%
\bibitem [{\citenamefont {Peters}\ \emph {et~al.}(2019)\citenamefont {Peters},
  \citenamefont {van Loon}, \citenamefont {Rubtsov}, \citenamefont
  {Lichtenstein}, \citenamefont {Katsnelson},\ and\ \citenamefont
  {Stepanov}}]{PhysRevB.100.165128}%
  \BibitemOpen
  \bibfield  {author} {\bibinfo {author} {\bibfnamefont {L.}~\bibnamefont
  {Peters}}, \bibinfo {author} {\bibfnamefont {E.~G. C.~P.}\ \bibnamefont {van
  Loon}}, \bibinfo {author} {\bibfnamefont {A.~N.}\ \bibnamefont {Rubtsov}},
  \bibinfo {author} {\bibfnamefont {A.~I.}\ \bibnamefont {Lichtenstein}},
  \bibinfo {author} {\bibfnamefont {M.~I.}\ \bibnamefont {Katsnelson}}, \ and\
  \bibinfo {author} {\bibfnamefont {E.~A.}\ \bibnamefont {Stepanov}},\
  }\bibfield  {title} {\enquote {\bibinfo {title} {{Dual boson approach with
  instantaneous interaction}},}\ }\href {\doibase 10.1103/PhysRevB.100.165128}
  {\bibfield  {journal} {\bibinfo  {journal} {Phys. Rev. B}\ }\textbf {\bibinfo
  {volume} {100}},\ \bibinfo {pages} {165128} (\bibinfo {year}
  {2019})}\BibitemShut {NoStop}%
\bibitem [{\citenamefont {Galler}\ \emph {et~al.}(2017)\citenamefont {Galler},
  \citenamefont {Thunstr\"om}, \citenamefont {Gunacker}, \citenamefont
  {Tomczak},\ and\ \citenamefont {Held}}]{PhysRevB.95.115107}%
  \BibitemOpen
  \bibfield  {author} {\bibinfo {author} {\bibfnamefont {A.}~\bibnamefont
  {Galler}}, \bibinfo {author} {\bibfnamefont {P.}~\bibnamefont {Thunstr\"om}},
  \bibinfo {author} {\bibfnamefont {P.}~\bibnamefont {Gunacker}}, \bibinfo
  {author} {\bibfnamefont {J.~M.}\ \bibnamefont {Tomczak}}, \ and\ \bibinfo
  {author} {\bibfnamefont {K.}~\bibnamefont {Held}},\ }\bibfield  {title}
  {\enquote {\bibinfo {title} {{Ab initio dynamical vertex approximation}},}\
  }\href {\doibase 10.1103/PhysRevB.95.115107} {\bibfield  {journal} {\bibinfo
  {journal} {Phys. Rev. B}\ }\textbf {\bibinfo {volume} {95}},\ \bibinfo
  {pages} {115107} (\bibinfo {year} {2017})}\BibitemShut {NoStop}%
\bibitem [{\citenamefont {Galler}\ \emph {et~al.}(2018)\citenamefont {Galler},
  \citenamefont {Kaufmann}, \citenamefont {Gunacker}, \citenamefont {Pickem},
  \citenamefont {Thunstr\"om}, \citenamefont {Tomczak},\ and\ \citenamefont
  {Held}}]{doi:10.7566/JPSJ.87.041004}%
  \BibitemOpen
  \bibfield  {author} {\bibinfo {author} {\bibfnamefont {A.}~\bibnamefont
  {Galler}}, \bibinfo {author} {\bibfnamefont {J.}~\bibnamefont {Kaufmann}},
  \bibinfo {author} {\bibfnamefont {P.}~\bibnamefont {Gunacker}}, \bibinfo
  {author} {\bibfnamefont {M.}~\bibnamefont {Pickem}}, \bibinfo {author}
  {\bibfnamefont {P.}~\bibnamefont {Thunstr\"om}}, \bibinfo {author}
  {\bibfnamefont {J.~M.}\ \bibnamefont {Tomczak}}, \ and\ \bibinfo {author}
  {\bibfnamefont {K.}~\bibnamefont {Held}},\ }\bibfield  {title} {\enquote
  {\bibinfo {title} {{Towards ab initio Calculations with the Dynamical Vertex
  Approximation}},}\ }\href {\doibase 10.7566/JPSJ.87.041004} {\bibfield
  {journal} {\bibinfo  {journal} {J. Phys. Soc. Jpn.}\ }\textbf {\bibinfo
  {volume} {87}},\ \bibinfo {pages} {041004} (\bibinfo {year}
  {2018})}\BibitemShut {NoStop}%
\bibitem [{\citenamefont {Cao}\ \emph {et~al.}(2018)\citenamefont {Cao},
  \citenamefont {Ayral}, \citenamefont {Zhong}, \citenamefont {Parcollet},
  \citenamefont {Manske},\ and\ \citenamefont {Hansmann}}]{PhysRevB.97.155145}%
  \BibitemOpen
  \bibfield  {author} {\bibinfo {author} {\bibfnamefont {X.}~\bibnamefont
  {Cao}}, \bibinfo {author} {\bibfnamefont {T.}~\bibnamefont {Ayral}}, \bibinfo
  {author} {\bibfnamefont {Z.}~\bibnamefont {Zhong}}, \bibinfo {author}
  {\bibfnamefont {O.}~\bibnamefont {Parcollet}}, \bibinfo {author}
  {\bibfnamefont {D.}~\bibnamefont {Manske}}, \ and\ \bibinfo {author}
  {\bibfnamefont {P.}~\bibnamefont {Hansmann}},\ }\bibfield  {title} {\enquote
  {\bibinfo {title} {{Chiral $d$-wave superconductivity in a triangular surface
  lattice mediated by long-range interaction}},}\ }\href {\doibase
  10.1103/PhysRevB.97.155145} {\bibfield  {journal} {\bibinfo  {journal} {Phys.
  Rev. B}\ }\textbf {\bibinfo {volume} {97}},\ \bibinfo {pages} {155145}
  (\bibinfo {year} {2018})}\BibitemShut {NoStop}%
\bibitem [{\citenamefont {Vandelli}\ \emph {et~al.}(2022)\citenamefont
  {Vandelli}, \citenamefont {Kaufmann}, \citenamefont {El-Nabulsi},
  \citenamefont {Harkov}, \citenamefont {Lichtenstein},\ and\ \citenamefont
  {Stepanov}}]{arxiv.2204.06426}%
  \BibitemOpen
  \bibfield  {author} {\bibinfo {author} {\bibfnamefont {Matteo}\ \bibnamefont
  {Vandelli}}, \bibinfo {author} {\bibfnamefont {Josef}\ \bibnamefont
  {Kaufmann}}, \bibinfo {author} {\bibfnamefont {Mohammed}\ \bibnamefont
  {El-Nabulsi}}, \bibinfo {author} {\bibfnamefont {Viktor}\ \bibnamefont
  {Harkov}}, \bibinfo {author} {\bibfnamefont {Alexander~I.}\ \bibnamefont
  {Lichtenstein}}, \ and\ \bibinfo {author} {\bibfnamefont {Evgeny~A.}\
  \bibnamefont {Stepanov}},\ }\href {\doibase 10.48550/ARXIV.2204.06426}
  {\enquote {\bibinfo {title} {{Multi-band D-TRILEX approach to materials with
  strong electronic correlations}},}\ }\bibinfo {howpublished} {Preprint
  arXiv:2204.06426} (\bibinfo {year} {2022})\BibitemShut {NoStop}%
\bibitem [{\citenamefont {Rubtsov}(2018)}]{PhysRevE.97.052120}%
  \BibitemOpen
  \bibfield  {author} {\bibinfo {author} {\bibfnamefont {A.~N.}\ \bibnamefont
  {Rubtsov}},\ }\bibfield  {title} {\enquote {\bibinfo {title} {{Fluctuating
  local field method probed for a description of small classical correlated
  lattices}},}\ }\href {\doibase 10.1103/PhysRevE.97.052120} {\bibfield
  {journal} {\bibinfo  {journal} {Phys. Rev. E}\ }\textbf {\bibinfo {volume}
  {97}},\ \bibinfo {pages} {052120} (\bibinfo {year} {2018})}\BibitemShut
  {NoStop}%
\bibitem [{\citenamefont {Rubtsov}\ \emph {et~al.}(2020)\citenamefont
  {Rubtsov}, \citenamefont {Stepanov},\ and\ \citenamefont
  {Lichtenstein}}]{PhysRevB.102.224423}%
  \BibitemOpen
  \bibfield  {author} {\bibinfo {author} {\bibfnamefont {A.~N.}\ \bibnamefont
  {Rubtsov}}, \bibinfo {author} {\bibfnamefont {E.~A.}\ \bibnamefont
  {Stepanov}}, \ and\ \bibinfo {author} {\bibfnamefont {A.~I.}\ \bibnamefont
  {Lichtenstein}},\ }\bibfield  {title} {\enquote {\bibinfo {title}
  {{Collective magnetic fluctuations in Hubbard plaquettes captured by
  fluctuating local field method}},}\ }\href {\doibase
  10.1103/PhysRevB.102.224423} {\bibfield  {journal} {\bibinfo  {journal}
  {Phys. Rev. B}\ }\textbf {\bibinfo {volume} {102}},\ \bibinfo {pages}
  {224423} (\bibinfo {year} {2020})}\BibitemShut {NoStop}%
\bibitem [{\citenamefont {Lyakhova}\ \emph {et~al.}(2022)\citenamefont
  {Lyakhova}, \citenamefont {Stepanov},\ and\ \citenamefont
  {Rubtsov}}]{PhysRevB.105.035118}%
  \BibitemOpen
  \bibfield  {author} {\bibinfo {author} {\bibfnamefont {Y.~S.}\ \bibnamefont
  {Lyakhova}}, \bibinfo {author} {\bibfnamefont {E.~A.}\ \bibnamefont
  {Stepanov}}, \ and\ \bibinfo {author} {\bibfnamefont {A.~N.}\ \bibnamefont
  {Rubtsov}},\ }\bibfield  {title} {\enquote {\bibinfo {title} {{Fluctuating
  local field approach to free energy of one-dimensional molecules with strong
  collective electronic fluctuations}},}\ }\href {\doibase
  10.1103/PhysRevB.105.035118} {\bibfield  {journal} {\bibinfo  {journal}
  {Phys. Rev. B}\ }\textbf {\bibinfo {volume} {105}},\ \bibinfo {pages}
  {035118} (\bibinfo {year} {2022})}\BibitemShut {NoStop}%
\bibitem [{\citenamefont {Lyakhova}\ and\ \citenamefont
  {Rubtsov}(2022)}]{s10948-022-06303-8}%
  \BibitemOpen
  \bibfield  {author} {\bibinfo {author} {\bibfnamefont {Y.~S.}\ \bibnamefont
  {Lyakhova}}\ and\ \bibinfo {author} {\bibfnamefont {A.~N.}\ \bibnamefont
  {Rubtsov}},\ }\bibfield  {title} {\enquote {\bibinfo {title} {{Fluctuating
  local field approach to the description of lattice models in the strong
  coupling regime}},}\ }\href {\doibase 10.1007/s10948-022-06303-8} {\bibfield
  {journal} {\bibinfo  {journal} {J. Supercond. Nov. Magn.}\ }\textbf {\bibinfo
  {volume} {35}},\ \bibinfo {pages} {2169–2173} (\bibinfo {year}
  {2022})}\BibitemShut {NoStop}%
\bibitem [{\citenamefont {Ayral}\ \emph {et~al.}(2017)\citenamefont {Ayral},
  \citenamefont {Biermann}, \citenamefont {Werner},\ and\ \citenamefont
  {Boehnke}}]{PhysRevB.95.245130}%
  \BibitemOpen
  \bibfield  {author} {\bibinfo {author} {\bibfnamefont {T.}~\bibnamefont
  {Ayral}}, \bibinfo {author} {\bibfnamefont {S.}~\bibnamefont {Biermann}},
  \bibinfo {author} {\bibfnamefont {P.}~\bibnamefont {Werner}}, \ and\ \bibinfo
  {author} {\bibfnamefont {L.}~\bibnamefont {Boehnke}},\ }\bibfield  {title}
  {\enquote {\bibinfo {title} {{Influence of Fock exchange in combined
  many-body perturbation and dynamical mean field theory}},}\ }\href {\doibase
  10.1103/PhysRevB.95.245130} {\bibfield  {journal} {\bibinfo  {journal} {Phys.
  Rev. B}\ }\textbf {\bibinfo {volume} {95}},\ \bibinfo {pages} {245130}
  (\bibinfo {year} {2017})}\BibitemShut {NoStop}%
\bibitem [{\citenamefont {Peierls}(1938)}]{PhysRev.54.918}%
  \BibitemOpen
  \bibfield  {author} {\bibinfo {author} {\bibfnamefont {R.}~\bibnamefont
  {Peierls}},\ }\bibfield  {title} {\enquote {\bibinfo {title} {{On a Minimum
  Property of the Free Energy}},}\ }\href {\doibase 10.1103/PhysRev.54.918}
  {\bibfield  {journal} {\bibinfo  {journal} {Phys. Rev.}\ }\textbf {\bibinfo
  {volume} {54}},\ \bibinfo {pages} {918--919} (\bibinfo {year}
  {1938})}\BibitemShut {NoStop}%
\bibitem [{\citenamefont {Bogolyubov}(1958)}]{Bogolyubov:1958zv}%
  \BibitemOpen
  \bibfield  {author} {\bibinfo {author} {\bibfnamefont {N.~N.}\ \bibnamefont
  {Bogolyubov}},\ }\bibfield  {title} {\enquote {\bibinfo {title} {{On a
  variational principle in the many-body problem}},}\ }\href@noop {} {\bibfield
   {journal} {\bibinfo  {journal} {Sov. Phys. Dokl.}\ }\textbf {\bibinfo
  {volume} {3}},\ \bibinfo {pages} {292--294} (\bibinfo {year}
  {1958})}\BibitemShut {NoStop}%
\bibitem [{\citenamefont {Feynman}(1972)}]{feynman1972}%
  \BibitemOpen
  \bibfield  {author} {\bibinfo {author} {\bibfnamefont {R.~P.}\ \bibnamefont
  {Feynman}},\ }\href@noop {} {\emph {\bibinfo {title} {Statistical mechanics:
  A set of lectures}}}\ (\bibinfo  {publisher} {Reading, Mass:
  Benjamin/Cummings},\ \bibinfo {year} {1972})\BibitemShut {NoStop}%
\bibitem [{\citenamefont {Jaeckel}\ and\ \citenamefont
  {Wetterich}(2003)}]{PhysRevD.68.025020}%
  \BibitemOpen
  \bibfield  {author} {\bibinfo {author} {\bibfnamefont {J.}~\bibnamefont
  {Jaeckel}}\ and\ \bibinfo {author} {\bibfnamefont {C.}~\bibnamefont
  {Wetterich}},\ }\bibfield  {title} {\enquote {\bibinfo {title} {{Flow
  equations without mean field ambiguity}},}\ }\href {\doibase
  10.1103/PhysRevD.68.025020} {\bibfield  {journal} {\bibinfo  {journal} {Phys.
  Rev. D}\ }\textbf {\bibinfo {volume} {68}},\ \bibinfo {pages} {025020}
  (\bibinfo {year} {2003})}\BibitemShut {NoStop}%
\bibitem [{\citenamefont {Baier}\ \emph {et~al.}(2004)\citenamefont {Baier},
  \citenamefont {Bick},\ and\ \citenamefont {Wetterich}}]{PhysRevB.70.125111}%
  \BibitemOpen
  \bibfield  {author} {\bibinfo {author} {\bibfnamefont {T.}~\bibnamefont
  {Baier}}, \bibinfo {author} {\bibfnamefont {E.}~\bibnamefont {Bick}}, \ and\
  \bibinfo {author} {\bibfnamefont {C.}~\bibnamefont {Wetterich}},\ }\bibfield
  {title} {\enquote {\bibinfo {title} {{Temperature dependence of
  antiferromagnetic order in the Hubbard model}},}\ }\href {\doibase
  10.1103/PhysRevB.70.125111} {\bibfield  {journal} {\bibinfo  {journal} {Phys.
  Rev. B}\ }\textbf {\bibinfo {volume} {70}},\ \bibinfo {pages} {125111}
  (\bibinfo {year} {2004})}\BibitemShut {NoStop}%
\bibitem [{\citenamefont {Jaeckel}(2002)}]{jaeckel2002understanding}%
  \BibitemOpen
  \bibfield  {author} {\bibinfo {author} {\bibfnamefont {J.}~\bibnamefont
  {Jaeckel}},\ }\href {\doibase 10.48550/ARXIV.HEP-PH/0205154} {\enquote
  {\bibinfo {title} {{Understanding the Fierz Ambiguity of Partially Bosonized
  Theories}},}\ }\bibinfo {howpublished} {Preprint arXiv:0205154} (\bibinfo
  {year} {2002})\BibitemShut {NoStop}%
\bibitem [{\citenamefont {Stepanov}\ \emph
  {et~al.}(2019{\natexlab{b}})\citenamefont {Stepanov}, \citenamefont
  {Harkov},\ and\ \citenamefont {Lichtenstein}}]{PhysRevB.100.205115}%
  \BibitemOpen
  \bibfield  {author} {\bibinfo {author} {\bibfnamefont {E.~A.}\ \bibnamefont
  {Stepanov}}, \bibinfo {author} {\bibfnamefont {V.}~\bibnamefont {Harkov}}, \
  and\ \bibinfo {author} {\bibfnamefont {A.~I.}\ \bibnamefont {Lichtenstein}},\
  }\bibfield  {title} {\enquote {\bibinfo {title} {{Consistent partial
  bosonization of the extended Hubbard model}},}\ }\href {\doibase
  10.1103/PhysRevB.100.205115} {\bibfield  {journal} {\bibinfo  {journal}
  {Phys. Rev. B}\ }\textbf {\bibinfo {volume} {100}},\ \bibinfo {pages}
  {205115} (\bibinfo {year} {2019}{\natexlab{b}})}\BibitemShut {NoStop}%
\bibitem [{\citenamefont {Harkov}\ \emph {et~al.}(2021)\citenamefont {Harkov},
  \citenamefont {Vandelli}, \citenamefont {Brener}, \citenamefont
  {Lichtenstein},\ and\ \citenamefont {Stepanov}}]{PhysRevB.103.245123}%
  \BibitemOpen
  \bibfield  {author} {\bibinfo {author} {\bibfnamefont {V.}~\bibnamefont
  {Harkov}}, \bibinfo {author} {\bibfnamefont {M.}~\bibnamefont {Vandelli}},
  \bibinfo {author} {\bibfnamefont {S.}~\bibnamefont {Brener}}, \bibinfo
  {author} {\bibfnamefont {A.~I.}\ \bibnamefont {Lichtenstein}}, \ and\
  \bibinfo {author} {\bibfnamefont {E.~A.}\ \bibnamefont {Stepanov}},\
  }\bibfield  {title} {\enquote {\bibinfo {title} {{Impact of partially
  bosonized collective fluctuations on electronic degrees of freedom}},}\
  }\href {\doibase 10.1103/PhysRevB.103.245123} {\bibfield  {journal} {\bibinfo
   {journal} {Phys. Rev. B}\ }\textbf {\bibinfo {volume} {103}},\ \bibinfo
  {pages} {245123} (\bibinfo {year} {2021})}\BibitemShut {NoStop}%
\bibitem [{\citenamefont {\ifmmode~\check{S}\else \v{S}\fi{}imkovic}\ \emph
  {et~al.}(2020)\citenamefont {\ifmmode~\check{S}\else \v{S}\fi{}imkovic},
  \citenamefont {LeBlanc}, \citenamefont {Kim}, \citenamefont {Deng},
  \citenamefont {Prokof'ev}, \citenamefont {Svistunov},\ and\ \citenamefont
  {Kozik}}]{PhysRevLett.124.017003}%
  \BibitemOpen
  \bibfield  {author} {\bibinfo {author} {\bibfnamefont {F.}~\bibnamefont
  {\ifmmode~\check{S}\else \v{S}\fi{}imkovic}}, \bibinfo {author}
  {\bibfnamefont {J.~P.~F.}\ \bibnamefont {LeBlanc}}, \bibinfo {author}
  {\bibfnamefont {A.~J.}\ \bibnamefont {Kim}}, \bibinfo {author} {\bibfnamefont
  {Y.}~\bibnamefont {Deng}}, \bibinfo {author} {\bibfnamefont {N.~V.}\
  \bibnamefont {Prokof'ev}}, \bibinfo {author} {\bibfnamefont {B.~V.}\
  \bibnamefont {Svistunov}}, \ and\ \bibinfo {author} {\bibfnamefont
  {E.}~\bibnamefont {Kozik}},\ }\bibfield  {title} {\enquote {\bibinfo {title}
  {{Extended Crossover from a Fermi Liquid to a Quasiantiferromagnet in the
  Half-Filled 2D Hubbard Model}},}\ }\href {\doibase
  10.1103/PhysRevLett.124.017003} {\bibfield  {journal} {\bibinfo  {journal}
  {Phys. Rev. Lett.}\ }\textbf {\bibinfo {volume} {124}},\ \bibinfo {pages}
  {017003} (\bibinfo {year} {2020})}\BibitemShut {NoStop}%
\end{thebibliography}%

\end{document}